\documentclass[prd,aps,twocolumn,showpacs,amsmath,amssymb,superscriptaddress,floatfix,nofootinbib]{revtex4}
\usepackage[colorlinks=true, citecolor=blue, linkcolor=blue, urlcolor=blue]{hyperref}
\usepackage{graphicx}

\newcommand{\ds}{\displaystyle}
\newcommand{\DS}[1]{/\!\!\!#1}
\allowdisplaybreaks
\date{\today}

\begin{document}

\title{Branching fractions and polarizations of $D\to V \ell \nu_\ell$ within QCD light-cone sum rule}

\author{Hai-Bing Fu}
\address{Institute of Particle Physics $\&$ Department of Physics, Guizhou Minzu University, Guiyang 550025, P.R. China}
\author{Wei Cheng\footnote{Corresponding author}}
\email{chengwei@itp.ac.cn}
\address{CAS Key Laboratory of Theoretical Physics, Institute of Theoretical Physics, Chinese Academy of Sciences,
Beijing 100190, P.R. China}
\address{Institute of Theoretical Physics, Chinese Academy of Sciences,  P. O. Box 2735, Beijing 100190,  P.R. China}
\author{Long Zeng}
\address{Institute of Particle Physics $\&$ Department of Physics, Guizhou Minzu University, Guiyang 550025, P.R. China}
\author{Dan-Dan Hu}
\address{Institute of Particle Physics $\&$ Department of Physics, Guizhou Minzu University, Guiyang 550025, P.R. China}
\author{Tao Zhong}
\address{Institute of Particle Physics $\&$ Department of Physics, Guizhou Minzu University, Guiyang 550025, P.R. China}

\begin{abstract}
In this paper, we make a detailed study about the $D\to V$ helicity form factors (HFFs) within the framework of QCD light-cone sum rule (LCSR) up to twist-4 accuracy. After extrapolating the LCSR predictions of HFFs to the whole physical $q^2$-region, we get the longitudinal, transverse and total  $|V_{cq}|$-independent decay widths of semileptonic decay $D\to V\ell^+\nu_\ell$. Meanwhile, the branching fractions of these decays are also obtained by using the $D^0(D^+)$-meson lifetime, which agree well with the BES-III results within errors. As a further step, we also investigate the differential and mean predictions for charged lepton (vector meson) polarization in the final state $P_{\rm L,T}^\ell$ ($F_{\rm L,T}^\ell$), the forward-backward asymmetry ${\cal A}_{\rm FB}^\ell$, and the lepton-side convexity parameters ${\cal C}_{\rm F}^\ell$. Our predictions are consistent with Covariant Confining Quark Model results within the errors. Thus, we think the LCSR approach for HFFs is applicable for dealing with the $D$-meson semileptonic decays.
\end{abstract}

\pacs{13.25.Hw, 11.55.Hx, 12.38.Aw, 14.40.Be}
\maketitle

\section{Introduction}
Semileptonic decays of charm mesons to vector mesons are a significant component in deeper understanding the standard model (SM) in post-Higgs era. Those decays are not only directly related to CKM matrix elements providing a window to study the CP-violation problem~\cite{Charles:2004jd, Poluektov:2010wz, Bona:2016bno, Abe:2018wpn, Aaij:2019kcg, Funatsu:2019fry, Resmi:2019ajp, Belfatto:2019swo}, but also contain the flavour-changing neutral current (FCNC) processes, which are sensitive to new physics (NP) due to it occurs at least one loop level in SM~\cite{Drobnak:2008br, Botella:2014ska, Kim:2015zla, Melikhov:2019nkv, Oyulmaz:2019jqr}. For the $D$-meson semileptonic decay processes, BES-III~\cite{Ablikim:2015gyp, Ablikim:2015wel, Ablikim:2016xny} and CLEO~\cite{Coan:2005iu, Naik:2009tk, Martin:2011rd,Briere:2010zc,Onyisi:2013bjt,CLEO:2011ab} Collaborations have report the branching fraction and form factors at large recoil point. Recently, BES-III collaboration measure the $D \to \rho\ell\nu_\ell$ branching fraction in 2019, i.e. $\mathcal{B}({D^0 \to \rho^- e^+ \nu_e})=(1.445\pm0.058_{\rm stat}\pm0.039_{\rm syst})\times10^{-3}$ and $\mathcal{B}({D^+ \to \rho^0 e^+ \nu_e})= (1.860\pm0.070_{\rm stat}\pm0.061_{\rm syst}) \times10^{-3}$~\cite{Ablikim:2018qzz}; meanwhile they also present an improved measurement for the ${D \to K^*(892)^-e^+\nu_e}$ branching fraction, i.e. $\mathcal{B}({D^0\to K^*(892)^- e^+ \nu_e}) =(2.033\pm0.046_{\rm stat}\pm0.047_{\rm syst}) \times10^{-2}$~\cite{Ablikim:2018lmn} in the same year. It is worth noting that BES-III recently published their first measurements for the $D^+\to \omega \mu^+\nu_\mu$ branching fraction in 2020, i.e. $\mathcal{B}(D^+\to\omega\mu^+\nu_\mu)=(17.7\pm1.8_{\rm stat}\pm1.1_{\rm syst})\times10^{-4}$, which is realized by applying an $e^+e^-$ collision data sample corresponding to an integrated luminosity of $2.93~ {\rm fb}^{-1}$ collected with the BES-III detector at a center-of-mass energy of $3.773~\rm{GeV}$~\cite{BES-III:2020dbj}.

To further understanding of the $D \to V$ semileptonic decay processes, the complete angular distribution and polarization informations should be investigated, such as the longitudinal and transverse polarizations of the final charged lepton $P_{\rm L,T}^\ell(q^2)$ and the final vector meson $F_{\rm L,T}^\ell(q^2)$, the forward-backward asymmetry $\mathcal{A}_{\rm{FB}}^\ell(q^2)$, and the lepton-side convexity parameter ${\cal C}_{\rm F}^\ell(q^2)$. On the one hand, the little theoretical research on those polarization information and even less experimental ones, but one can find numerous experimental and theoretical studies for that of the $B$-meson~\cite{Grossman:2000rk, Gronau:2001ng, Kruger:2005ep, Kou:2010kn, Wehle:2016gfb, Cheng:2017bzz, Fu:2018hko,Zhou:2019jny, Wang:2019wee, Martin:2019xcc}. On the other hand, QCD light-cone sum rules (LCSR) is an effective method to study the heavy meson to light meson decay processes that are actually an FCNC process of the heavy quark to light quark transition. The $D \to V$ decay correspond to the transition of heavy quark to light quark $c \to q~(u,d,s)$, and the mass of charm-quark is the order magnitude of GeV. Which indicate LCSR is applicable in studying the $D \to V$ decay process. Therefore, in this paper, we will study those observations for the semileptonic $D \to V \ell^+\nu_\ell$ decays within the framework of LCSR.

The LCSR is an important method in dealing with the semileptonic decays~\cite{Braun:1997kw, Huang:1998gp, Huang:2001xb, Wu:2007vi,Momeni:2016kjz,Wang:2017jow,Cheng:2017fkw,Fu:2018yin,Gao:2019lta,Momeni:2019uag}. Its main strategy is to construct an analytic heavy-to-light correlator function in the whole $q^2$-region and then make an operator product expansion (OPE) and a hadron expression for it in the spacelike and timelike region respectively, finally combine the results achieved in those two ways to get the form factors with the help of Borel transformation. Both of transition form factors (TFFs) and helicity form factors (HFFs) contain the information of meson semileptonic decays independently, because they can describe the non-perturbative hadronic matrix elements of meson semileptonic decays independently. One can decompose the hadronic matrix elements by applying momentum of initial and final meson states to obtain TFFs~\cite{Balitsky:1989ry, Chernyak:1990ag, Belyaev:1993wp, Ball:1997rj, Ball:2004rg, Fajfer:2005ug, Wang:2008ci,Faustov:2019mqr}, which will lead to a mix of longitudinal and transverse polarization of the meson among those TFFs. Thus TFFs cannot express the polarization information of meson decay accurately.

The HFFs opened a new avenue to deal with those matrix elements~\cite{Bharucha:2010im,Cheng:2018ouz,Fu:2020uzy}. HFFs decompose it by applying the off-shell $W$-boson polarization vectors, which brings a good polarization property, i.e. researching on tracking polarization. Specifically, the longitudinal and transverse decay information can be completely separated, which is very useful for probing the longitudinal and transverse polarization separately. Specially, both of the decay width of $D$-meson longitudinal and transverse components contain the usual TFF ($A_1(q^2)$), which means $D$-meson longitudinal and transverse components are related to each other. However, in the case of HFFs, the decay widths of $D$-meson longitudinal (transverse) component related to $\mathcal{D}_{V,0}(q^2)$ ($\mathcal{D}_{V,1}(q^2)$, $\mathcal{D}_{V,2}(q^2)$) HFF, which means the decay widths of $D$-meson longitudinal and transverse component are separated completely and have no influence on each other. In addition, it also enjoys some other advantages, such as dispersive bounds on the HFF parametrization and direct relations between the HFFs and the spin-parity quantum numbers. For more detailed discussion, one can refer to the literatures~\cite{Bharucha:2010im,Cheng:2018ouz,Fu:2020uzy}.

\begin{table}[t]
\centering
\caption{Resonance masses of quantum number $J^P$ as indicated necessary for the parameterisation of $D\to V$ HFFs $\mathcal{D}_{V,\sigma}$~\cite{Tanabashi:2018oca,Leljak:2019eyw} with $\sigma = 0,1,2,t$ respectively.}\label{tab:mRi}
\renewcommand{\arraystretch}{1.5}
\addtolength{\arraycolsep}{6pt}
\begin{tabular}{l c c }
\hline
~~$F_i$~~                 &~~$J^P$~~ & ~~$m_{R,i}/{\rm GeV}$~~\\ \hline
$\mathcal{D}_{V,t}$       & $0^-$    & $1.864$  \\
$\mathcal{D}_{V,0;2}$     & $1^+$    & $2.420$  \\
$\mathcal{D}_{V,1}$       & $1^-$    & $2.007$\\ \hline
\end{tabular}
\addtolength{\arraycolsep}{-6pt}
\end{table}

The remaining parts of the paper are organized as follows. In Sec.~\ref{sec:II}, we introduce the physical observable for $D \to V(\rho,\omega,K^*)\ell^+\nu_\ell$ semileptonic decay processes, and calculate the HFFs within the LCSR approach. In Sec.~\ref{sec:III}, after fixing the hadron input parameters for HFFs and extrapolating those HFFs to the whole $q^2$-region with Simplified Series Expansion (SSE), we then apply it to investigate the $D \to V $ semileptonic decay observable, such as decay width, branching fraction and polarizations, and also compare our results with available experimental and other theoretical predictions. Finally, we briefly summarize in Sec.~\ref{sec:IV}.

\section{Calculation Technology}\label{sec:II}

\subsection{$D\to V \ell^+\nu_\ell$ semileptonic decays}

\begin{figure}[t]
\begin{center}
\includegraphics[width=0.40\textwidth]{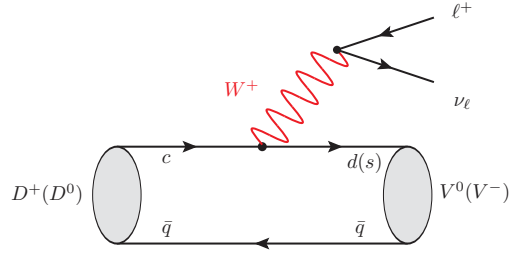}
\end{center}
\caption{Typical diagram for the $D \to V \ell^+\nu_\ell$ semileptonic decay, where $q=u,d$ and $V=\rho,\omega$,$K^*$-mesons.}
\label{Fig:DtoV}
\end{figure}

The $D \to V \ell^+\nu_\ell$ semileptonic decay process are displayed in Fig.~\ref{Fig:DtoV}. The corresponding invariant matrix element can be expressed as follows:
\begin{align}
\mathcal M(D\to V \ell^+ \nu_{\ell}) = \frac{G_F}{\sqrt{2}} V^\ast_{cq} H^\mu L_\mu,
\end{align}
where fermi constant $G_F=1.166\times10^{-5}~{\rm GeV}^{-2}$, leptonic current $L_\mu = \bar{\nu}_\ell \gamma_\mu(1-\gamma_5) \ell$ and the hadron matric element $H^\mu=\langle V |V^\mu-A^\mu |D \rangle$ with flavor-changing vector current $V^\mu=\bar{q}\gamma^\mu c$ and axial-vector current $A^\mu=\bar{q}\gamma^\mu\gamma_5c$.

To get accurate polarization properties of the semileptonic decay $D \to V \ell^+\nu_\ell$, one can decompose the hadron matric element $H^\mu$ into the HFFs by the off-shell $W$-boson polarization vectors. Specifically speaking,
\begin{equation}
{\cal D}_{V,\sigma}(q^2)= \sqrt{\frac{q^2}{\lambda}}\sum_{\alpha=0,\pm,t}
{\epsilon_\sigma^{*\mu}(q)}\langle V(\tilde p,\tilde\epsilon_\alpha)|\bar q \gamma_\mu(1-\gamma^5) c |D(p)\rangle,\label{eq:BVdef}
\end{equation}
where the standard kinematic function is $\lambda=(t_- - q^2)(t_+ - q^2)$ with $t_\pm = (m_D\pm m_V)^2$ and $\epsilon_\sigma ^{*\mu}(q)$ represent transverse ($\sigma=\pm$), longitudinal ($\sigma=0$) or time-like ($\sigma=t$) polarization vectors. For the convenience of polarization research, two HFFs, ${\cal D}_{V,(1,2)}(q^2)$, are defined by a linear combinations of the transverse helicity projection vector, i.e. $\epsilon_{(1,2)}(q)=[\epsilon_-(q) \mp \epsilon_+(q)]/\sqrt{2}$, which will be discussed in the following section.

The polar angle differential decay distribution in the momentum transfer squared, defined by the angle between $\vec q = \vec p_D - \vec p_V$ and the three-momentum of the charged lepton in the rest frame, can be written as follows,
\begin{align}
\frac{d^2\Gamma}{dq^2d\cos\theta} &= \frac{|{\bf p_V}|v}{(2\pi)^3 32 m_D^2} \sum_{\rm pol}|{\cal M}|^2,
\end{align}
where $|{\bf p_V}|=\lambda^{1/2}/(2m_D^2)$, $v = (1 - {m_\ell^2}/{q^2})$, and the covariant contraction $\sum_{\rm pol}|{\cal M}|^2$ can be converted to a sum of bilinear products of hadronic HFFs and leptonic helicity amplitude by applying the completeness relation to the polarization four-vectors of the process. So the total differential decay width of $D\to V\ell\nu_\ell$ can be expressed as,
\begin{align}
\frac{1}{|V_{\rm cq}|^2}\frac{d\Gamma}{dq^2} & = {\cal G} \lambda^{3/2} v^2 \bigg[(1+ \delta_\ell)\sum {\cal D}_{V,i}^2(q^2) + 3 \delta_\ell {\cal D}_{V,t}^2(q^2)\bigg]  \nonumber\\
&= {\cal G} \lambda^{3/2} v^2 {\cal D}_{\rm tot}^2(q^2), \label{difftot}
\end{align}
with $\delta_\ell = m_\ell^2/(2q^2)$, ${\cal G}={G_F^2}/{(192\pi^3 m_D^3)}$ and variable $i = 0,1,2$. The detailed expression reads,
\begin{align}
\sum {\cal D}_{V,i}^2 = {\cal D}_{V,0}^2 + {\cal D}_{V,1}^2 +{\cal D}_{V,2}^2.
\end{align}

As we know that the total decay width can be separated into longitudinal and transverse parts, i.e. $\Gamma=\Gamma^{\rm L}+\Gamma^{\rm T}$. The decay width for the vector meson longitudinal component $\Gamma^{\rm L}$ is defined as
\begin{align}
\Gamma^{\rm L} (q^2)= {\cal G} |V_{cq}|^2 \int_0^{q^2_{\rm max}} dq^2 \lambda(q^2)^{3/2} {\cal D}_{V,0}^2(q^2)\label{diffL}
\end{align}
and the decay width for the vector meson transverse component $\Gamma^{\rm T}$ is defined as
\begin{align}
\Gamma^{\rm T} (q^2)= {\cal G} |V_{cq}|^2 \int_0^{q^2_{\rm max}} dq^2 \lambda(q^2)^{3/2}\sum {\cal D}_{V,j} ^2,\label{diffT}
\end{align}
where the variable $j$ is the summation index, and represented $j = 1,2$.

For the polarization properties of the semileptonic decay $D \to V \ell^+\nu_\ell$, one can study the longitudinal and transverse polarization firstly. Specifically, with the help of HFFs, the longitudinal $P_{\rm L}^\ell $ and transverse $P_{\rm T}^\ell$ polarization of the charged lepton in the final state and the longitudinal $F_{\rm L}^\ell$ and transverse $F_{\rm T}^\ell$ polarization fractions of the vector meson in the final state are given by:
\begin{eqnarray}
P_{\rm L}^\ell &=& 1 - \frac{2\sum {\cal D}_{V,i} ^2}{{\cal D}_{\rm tot}}~,\nonumber\\
P_{\rm T}^\ell &=& \frac{3\pi\sqrt{\delta_\ell}}{2\sqrt 2} \frac{{\cal D}_{V,1} {\cal D}_{V,2}  - {\cal D}_{V,0} {\cal D}_{V,t}}{{\cal D}_{\rm tot}}~,\nonumber\\
F_{\rm L}^\ell &=& \frac{{3{\cal D}_{V,t} ^2\delta_\ell+ (1 + \delta_\ell){\cal D}_{V,0} ^2}}{{\cal D}_{\rm tot}}~,\nonumber\\
F_{\rm T}^\ell &=& \frac{{(1 + \delta_\ell)({\cal D}_{V,1} ^2+{\cal D}_{V,2} ^2)}}{{\cal D}_{\rm tot}}.
\label{Eq:PLTFLT}
\end{eqnarray}
And the forward-backward asymmetry ${\cal A}_{\rm FB}^\ell$ can be written as,
\begin{align}
{\cal A}_{\rm FB}^\ell&=\frac{d\Gamma(F)-d\Gamma(B)}{d\Gamma(F)+d\Gamma(B)}
\nonumber\\
&= \frac{ \ds\int_0^1 d\!\cos\theta\, d\Gamma/d\!\cos\theta
      -\int_{-1}^0 d\!\cos\theta\, d\Gamma/d\!\cos\theta }
     { \ds\int_0^1 d\!\cos\theta\, d\Gamma/d\!\cos\theta
      +\int_{-1}^0 d\!\cos\theta\, d\Gamma/d\!\cos\theta}
\nonumber\\[0.5ex]
&=\frac32 \frac{2\delta_\ell {\cal D}_{V,0}{\cal D}_{V,t} - {\cal D}_{V,1}{\cal D}_{V,2} }{(1+\delta_\ell)\sum{\cal D}_{V,i}^2 + \delta_\ell{\cal D}_{V,t} ^2}~.
\label{eq:AFB}
\end{align}
The lepton-side ${\cal C}_{\rm F}^\ell$ convexity parameters has the form,
\begin{align}
{\cal C}_{\rm F}^\ell &= -\frac34\frac{(2{\cal D}_{V,0}^2 - {\cal D}_{V,1}^2- {\cal D}_{V,2} ^2)(1- 2\delta_\ell)}{{\cal D}_{\rm tot}}~.
\label{eq:CFl}
\end{align}
In order to make a comparison with other approaches, we take the same approach as the paper~\cite{Ivanov:2019nqd} to deal with the $q^2$ average of those observable. Specifically, if an observable $A$ has the form $A={{\cal D}_x}/{{\cal D}_y}$, one can multiply both the numerator and denominator the phase-space factor and then integrate the two separately. The detailed expression can be written as:
\begin{align}
\langle A \rangle = \dfrac{\ds\int C(q^2){\cal D}_x dq^2}{\ds\int C(q^2){\cal D}_y dq^2}~,
\label{eq:Aob}
\end{align}
with $q^2$ dependence phase-space factor
\begin{align}
C(q^2) = |{\bf p_V}|\frac{(q^2 - m^2_\ell )^2}{q^2}~.
\end{align}

\subsection{The $D \to V$ HFFs}
To derive LCSRs for the four HFFs i.e. ${\cal D}_{V,\sigma}(q^2)$ with $ \sigma = 0,1,2,t$, we first structure a two-point correlation function according to LCSR strategy, as follows:
\begin{align}
\Pi_\sigma(p,q)&=i \sqrt{\frac{q^2}{\lambda}} {\epsilon_\sigma^{*\mu}(q)}  \nonumber\\
&  \times \int d^4 x e^{iq\cdot x}\langle V(\tilde p,\tilde\epsilon)|T\{j_{V,\mu}(x),j_D^\dag (0)\}|0\rangle,\label{correlators}
\end{align}
where the hadron vector and pseudoscalar current are $j_{V,\mu}(x) = \bar {q}(x){\gamma_\mu} c(x)$ and $j_D^\dag (0)=\bar c(0)i \gamma_5 u(0)$ respectively. Here $T$ is the product of the current operator.

In the timelike $q^2$-region, after inserting the complete intermediate states that have the same quantum numbers $J^P=0^-$ with the current operator $\bar c i \gamma_5 u$ into the hadron current of the correlation function, and further isolating the pole term of the lowest pseudoscalar $D$-meson, the correlation function can be read off:
\begin{align}
\Pi_\sigma^{\rm H}(p,q)  &= \sqrt{\frac{q^2}{\lambda}} \bigg[ \frac{\ds{\epsilon_\sigma^{*\mu}(q)}\langle V|\bar{q}\gamma _\mu c|D\rangle \langle D|\bar c i \gamma_5 u|0\rangle} {m_D^2-(p+q)^2} \nonumber\\[1ex]
&+\sum\limits_{\rm H} \frac{{\epsilon_\sigma^{*\mu}(q)}\langle V|\bar{q}\gamma _\mu c|D^{\rm H}\rangle \langle D^{\rm H}|\bar{q}i\gamma_5 u|0\rangle}{m_{D^{\rm H}}^2-(p+q)^2}\bigg],
\end{align}
where $\langle D|\bar{c}i\gamma_5 u|0\rangle={m_D}^2f_D/m_c$. After replacing the sum of higher resonances and continuum states with the dispersion integrations, the hadronic representation of the correlator $\Pi^H_\sigma$ finally has the form:
\begin{align}
\Pi^{\rm H}_\sigma(q^2,(p+q)^2)&=\frac{m_D^2 f_D}{m_c [m_D^2-(p+q)^2]}\mathcal{D}_{\sigma}(q^2)
\nonumber\\[1ex]
&+\int_{s_0}^{\infty}\frac{\rho^{\rm H}_\sigma(s)}{s-(p+q)^2}ds + {\rm subtractions},
\end{align}
In the spacelike $q^2$-region, i.e. $(p+q)^2-m_c^2\ll 0$, and $q^2\ll m_c^2-{\cal O}(1{\rm GeV ^2})$, one needs to contract the c-quark operator by applying a propagator with the gluon field correction:
\begin{align}
\langle0|c_\alpha^i(x)&\bar c_\beta^j(0)|0\rangle  = -i\int \frac{d^4k}{(2\pi)^4}e^{-ik\cdot x}\bigg\{\delta^{ij}\frac{\DS k + m_c}{m_c^2-k^2} \nonumber\\[0.5ex]
&+g_s\int_0^1 dv G^{\mu\nu\alpha}(vx)\left(\frac{\lambda}{2}\right)^{ij}
\nonumber\\[0.5ex]
&\times\bigg[\frac{\DS k+m_c}{2(m_c^2 - k^2)^2}\sigma_{\mu\nu}+ \frac1{m_c^2-k^2}vx_\mu\gamma_\nu\bigg]\bigg\}_{\alpha\beta}.
\end{align}
For further OPE treatment, one needs the nonlocal matrix elements which are convoluted with the meson light-cone distribution amplitudes (LCDAs) of a growing twist:
\begin{widetext}
\begin{align}
\langle V (\tilde p,\tilde\epsilon)|\bar q_1(x)\sigma_{\mu \nu}q_2(0)|0\rangle &=- i f_V^\bot \int_0^1 du e^{iu(\tilde p\cdot x)}\bigg\{(\tilde\epsilon_\mu^*\tilde p_\nu  - \tilde\epsilon_\nu^*\tilde p_\mu)\bigg[\phi_{2;V}^\bot(u)+ \frac{m_V^2 x^2}{4} \phi_{4;V}^\bot(u)\bigg]
\nonumber\\
& +(\tilde p_\mu x_\nu - \tilde p_\nu x_\mu )\frac{\tilde\epsilon^*\cdot x}{(\tilde p\cdot x)^2} m_V^2 \left[\phi_{3;V}^\|(u) \,-\, \frac{1}{2}\phi_{2;V}^\bot(u) \,- \frac{1}{2} \psi_{4;V}^\bot(u)\right]\nonumber\\
& + \frac12 \left(\tilde\epsilon_\mu^*{x_\nu } - \tilde\epsilon_\nu^*x_\mu\right) \frac{m_V^2}{\tilde p\cdot x}
\left[ \psi_{4;V}^\bot(u)-\phi_{2;V}^\bot(u) \right] \bigg\}~,
\label{DA1}
\\[2ex]
\langle V (\tilde p,\tilde\epsilon )|\bar q_1(x) q_2(0)|0\rangle &= - \frac{i}{2} f_V^\bot\left(\tilde\epsilon^* \cdot x\right) m_V^2\int_0^1 du e^{iu(\tilde p\cdot x)} \psi_{3;V }^\parallel(u)~,
\label{DA2}\\[2ex]
\langle V (\tilde p,\tilde\epsilon)|\bar q_1(x)\gamma_\mu q_2(0)|0\rangle &= m_V^2 f_V^{\|}~\int_0^1~du ~e^{iu(\tilde p\cdot x)}~\bigg\{\frac{\tilde \epsilon^*\cdot x}{\tilde p \cdot x}\tilde p_\mu \bigg[\phi_{2;V}^\|(u) ~+~ \frac{m_V^2 x^2}{4} \phi_{4;V}^\|(u)\bigg]
\nonumber\\
& + \bigg(\tilde \epsilon_\mu^* -\tilde p_\mu \frac{\tilde \epsilon^*\cdot x}{p\cdot x} \bigg)\phi_{3;V}^\bot(u) -\frac12 m_V^2 x_\mu \frac{\tilde \epsilon^*\cdot x}{(\tilde p\cdot x)^2}\bigg[\psi_{4;V}^\|(u) +\phi_{2;V}^\|(u)
\nonumber\\
& - 2\phi_{3;V}^\bot(u)\bigg]\bigg\}~,
\label{DA3}
\\[2ex]
\langle V (\tilde p,\tilde\epsilon )|\bar q_1(x)\gamma_\mu\gamma_5 q_2(0)|0\rangle &= - \frac14 m_V f_V^\| \varepsilon^{\mu\nu\alpha\beta}\tilde\epsilon_\mu^* \tilde p_\alpha x_\beta \int_0^1 du e^{iu(\tilde p\cdot x)} \psi_{3;V}^\bot(u)~,
\label{DA4}
\end{align}
\end{widetext}
where $V = \rho, \omega, K^*$-mesons and $q_1=d(s)$ for $\rho,\omega,K^*$-mesons.

After replacing those hadronic matric elements and subtracting the contribution of the continuum spectrum using dispersion integration, one can finish the QCD representation calculation. In this paper, we will not consider the three-particle part due to its negligible contribution. Specifically, it is no more than $0.3\%$ of the total TFFs, and a more detailed analysis can be obtained from our previous study~\cite{Fu:2014pba}.

Moreover, one needs to equate the two types of representation of correlator and subtract the contributions from higher resonances and continuum states. With the help of Borel transformation, the LCSR for $D \to V$ HFFs can be finally read off:

\begin{widetext}
\begin{align}
{\cal D}_{V,0}(q^2) &  = \int_0^1 du e^{(m_D^2 - s)/M^2} ~\frac{m_c f_V^\bot{\cal F}}{2\sqrt\lambda  m_V m_D^2 f_D}~\bigg\{2{\cal S}\Theta (c(u,s_0)) \phi_{2;V}^\bot (u) - \frac{\lambda m_c m_V \tilde f_V}{u^2M^2}\widetilde \Theta (c(u,s_0))
\nonumber\\
&\times \Phi_{2;V}^\| (u) - (\tilde m_q  m_V{\tilde f_V} - m_V^2)\bigg[{\cal F}\Theta (c(u,s_0)) - \lambda \frac{1}{{uM^2}}\widetilde \Theta (c(u,s_0))\bigg] \psi_{3;V}^\| (u) ~ + m_c m_V
\nonumber\\
&\times \tilde f_V \bigg[\frac{\lambda}{u^2M^2} \widetilde\Theta(c(u,s_0)) \Phi_{3;V}^\bot (u)\,+\, {\cal F}\Theta (c(u,s_0))\phi_{3;V}^\bot (u)\bigg] + m_V^2{\cal S}\bigg[\frac{{\cal N}}{2u^3M^4}\widetilde{\widetilde \Theta} (c(u,s_0))
\nonumber\\
&- \frac{3}{{2u^2M^2}}\widetilde \Theta (c(u,s_0))\bigg] \phi_{4;V}^\bot (u) \,-\, \bigg[\frac{\lambda {\cal S}}{2u^3M^4}\widetilde{\widetilde \Theta} (c(u,s_0)) - m_V^2\frac{{{\cal S} - 4\lambda }}{u^2M^2}\bigg]
\widetilde\Theta (c(u,s_0)){I_L}(u)
\nonumber\\
& - \frac{{\lambda m_c^3m_V^3{\tilde f_V}}}{{{u^4}{M^6}}}\widetilde{\widetilde \Theta} (c(u,s_0))\Phi _{4;V}^ \bot (u) +  m_cm_V^3{\tilde f_V} \bigg[\frac{\lambda }{{u^2M^4}}\widetilde{\widetilde \Theta} (c(u,s_0)) + \frac{{\cal F}}{u^2M^2}\widetilde \Theta (c(u,s_0))\bigg]
\nonumber\\
&\times C_V(u) - m_V^2\bigg[ {\frac{3}{2}\Theta (c(u,s_0)) + \left( {\frac{{\cal N}}{u^2M^2} - \frac{\lambda}{{2uM^2{\cal F}}}} \right) \widetilde \Theta (c(u,s_0))} \bigg]{H_3}(u),
\label{eq:HV0}
\\
\nonumber\\
{\cal D}_{V,1}(q^2) & = \int_0^1 du e^{(m_D^2 - s)/M^2} \frac{m_c f_V^\bot \sqrt{2 q^2}} {2m_D^2 f_D} \bigg\{ \Theta (c(u,s_0))\phi _{2;V}^\bot (u) + m_V^2\bigg[\frac{{\cal N}}{u^3M^4}\widetilde{\widetilde\Theta} (c(u,s_0 )) + \frac{3}{u^2M^2}
\nonumber\\
&\times \widetilde \Theta (c(u,s_0 ))\bigg]\phi_{4;V}^ \bot (u) - \frac{{m_Vm_c{\tilde f_V}}}{{2u^2M^2}}\widetilde \Theta (c(u,s_0 ))\psi_{3;V}^\bot (u)\bigg\},
\label{eq:HV1}
\\
\nonumber\\
{\cal D}_{V,2}(q^2) &  = \int_0^1 du e^{(m_D^2 - s)/M^2} ~\frac{\sqrt {2q^2} m_c f_V^\bot}{2\sqrt \lambda m_D^2 f_D}~\bigg\{ {\cal E}\,\Theta (c(u,s_0))\phi_{2;V}^\bot(u) \,-\,2\,\Theta(c(u,s_0))\,(\tilde f_V m_V \tilde m_q
\nonumber\\
&- m_V^2)\,\psi_{3;V}^\|(u) \,+\, m_V^2 {\cal E}\bigg[\frac{{\cal N}}{u^3M^4}\widetilde{\widetilde \Theta} (c(u,s_0)) \,+\, \frac{3}{u^2M^2}\widetilde \Theta (c(u,s_0))\bigg]~\phi _{4;V}^ \bot (u)\,+\, \frac{2m_V^2}{u^2 M^2}
\nonumber\\
& \times {\cal E}\widetilde \Theta (c(u,s_0))I_L(u) \,- m_V^2\bigg[3\Theta (c(u,s_0)) \,+\, \frac{2{\cal N}}{u^2 M^2} \widetilde\Theta(c(u,s_0))\bigg]H_3(u) \,-\, 2m_c\tilde f_V m_V
\nonumber\\
& \times \Theta (c(u,s_0)) \phi_{3;V}^\bot (u) - \frac{2 m_c m_V^3 \tilde f_V}{u^2 M^2} \widetilde\Theta (c(u,s_0)) C_V(u)\bigg\},
\label{eq:HV2}
\\
\nonumber\\
{\cal D}_{V,t}(q^2)&  =  \int_0^1 du ~ e^{(m_D^2 - s)/M^2} ~\frac{m_c m_V f_V^\bot}{2m_V m_D^2 f_D} ~\bigg\{ u m_V \Theta (c(u,s_0))~\phi_{2;V}^\bot(u) \,-\, \frac{ m_c \tilde f_V\cal F}{u^2 M^2}~\widetilde\Theta(c(u,s_0))
\nonumber\\
&\times\Phi_{2;V}^\| (u) \,-\, (\tilde m_q \tilde f_V\,-\, m_V )\bigg[\Theta (c(u,s_0)) \,+\, \frac{{u{\cal F} + 2{q^2}}}{u^2 M^2}\widetilde \Theta (c(u,s_0))\bigg]\psi _{3;V}^\| (u) ~-\, m_c \tilde f_V
\nonumber\\
&\times \Theta (c(u,s_0))~\phi_{3;V}^\bot(u) ~+~ m_c \tilde f_V \frac{\cal F}{u^2 M^2}\widetilde\Theta(c(u,s_0))~\Phi_{3;V}^\bot (u) ~+~ m_V^3\bigg[\frac{{\cal N}}{u^2 M^4}\widetilde {\widetilde \Theta }(c(u,s_0)) \nonumber\\
&+\,\frac{3}{{u{M^2}}}\widetilde \Theta (c(u,s_0))\bigg]~\phi _{4;V}^ \bot (u) ~+~ m_c^3m_V^2 \tilde f_V ~\frac{{\cal F}}{u^4M^6}\,\widetilde {\widetilde {\widetilde \Theta }}(c(u,s_0))~\Phi_{4;V}^\| (u) \,-\,  m_V \bigg[\frac{{\cal E}}{{2u{M^2}}}
\nonumber\\
&\times \widetilde \Theta (c(u,s_0)) + \frac{3}{2}\Theta (c(u,s_0))\bigg]{H_3}(u) \,-\, m_V\bigg[\frac{{9{\cal F} - 2um_V^2 + 15{q^2}}}{u^2 M^2}\widetilde \Theta (c(u,s_0)) + \frac{{\cal W}}{u^3 M^4}
\nonumber\\
&\times\widetilde {\widetilde \Theta }(c(u,s_0))\bigg]{I_L}(u) + \frac{{m_c \tilde f_V }}{2}\bigg[\frac{{2m_V^2}}{u^2 M^2}\widetilde \Theta (c(u,s_0)) + \frac{{\cal S}}{u^3 M^4}\widetilde {\widetilde \Theta }(c(u,s_0))\bigg]C_V(u)\bigg\},\label{eq:HVt}
\end{align}
\end{widetext}
where ${\cal E} = m_D^2 + \xi m_V^2 - q^2$, ${\cal F} = m_D^2 - m_V^2 - q^2$, ${\cal N} = u m_D^2 - u \bar u m_V^2 + \bar u{q^2}$, ${\cal S} = 2m_V^2(um{D^2} - um_V^2 + (1 - \bar u){q^2})$, ${\cal W} = 2m_D^2[ - u\xi m_V^2 + {q^2}(1 + u + u\bar u)] + u\xi (m_D^4 + m_V^4) - 2{q^2}(1 + u)\bar um_V^2 - {q^4}(2 + u)$ and $s=[m_b^2-\bar u(q^2-u m_V^2)]/u$ with $\bar u=1-u$, $\xi = 2u-1$. The effective decay constant $\tilde f_V = f_V^\| / f_V^\bot$ and simplified distribution functions $\Phi_{2;V}^\|(u)$, $\Phi_{3;V}^\bot(u)$, $\Phi_{4;V}^\bot(u)$, $I_L(u)$ and $H_3(u)$ are defined as
\begin{align}
&\Phi_{2;V}^\|(u) = \int_0^u dv \phi_{2;V}^\|(u), \nonumber\\
&\Phi_{3;V}^\bot(u) = \int_0^u dv \phi_{3;V}^\bot(u), \nonumber\\
&\Phi_{4;V}^\bot(u) = \int_0^u dv \phi_{4;V}^\bot(u), \nonumber\\
&H_3(u) = \int_0^u dv \left[\psi_{4;V}^{\bot}(v)-\phi_{2;V}^\bot(v)\right], \nonumber\\
&I_L(u) = \int_0^u dv \int_0^v dw \left[\phi_{3;V}^\|(w) -\frac{1}{2} \phi_{2;V}^\bot(w) -\frac{1}{2} \psi_{4;V}^\bot(w)\right], \nonumber \\
&C_V(u) = \int_0^u {dv} \int_0^v {dw} \left[ {\psi _{4;V}^\| (w) + \phi_{2;V}^\| (w) - 2\phi _{3;V}^ \bot (w)} \right].
\end{align}
The $\Theta(c(u,s_0))$ is conventional step function, $\widetilde\Theta [c(u,s_0)]$ and $\widetilde{\widetilde\Theta}(c(u,s_0)]$ are defined as
\begin{align}
\int_0^1 \frac{du}{u^2 M^2} & e^{-s/M^2}\widetilde\Theta(c(u,s_0))f(u)\nonumber\\
&= \int_{u_0}^1\frac{du}{u^2 M^2} e^{-s)/M^2}f(u) + \delta(c(u_0,s_0)),
\label{Theta1}\\
\int_0^1 \frac{du}{2u^3 M^4} & e^{-s/M^2}\widetilde{\widetilde\Theta}(c(u,s_0))f(u)\nonumber\\
&= \int_{u_0}^1 \frac{du}{2u^3 M^4} e^{-s/M^2}f(u)+\Delta(c(u_0,s_0)), \label{Theta2}
\end{align}
where
\begin{align}
\delta(c(u,s_0))&= e^{-s_0/M^2}\frac{f(u_0)}{{\cal C}_0}, \nonumber\\
\Delta(c(u,s_0))&= e^{-s_0/M^2}\bigg[\frac{1}{2 u_0 M^2}\frac{f(u_0)} {{\cal C}_0} \left. -\frac{u_0^2}{2 {\cal C}_0} \frac{d}{du}\left( \frac{f(u)}{u{\cal C}} \right) \right|_{u = {u_0}}\bigg], \nonumber
\end{align}
${\mathcal C}_0 = m_b^2 + {u_0^2}m_V ^2 - {q^2}$ and $u_0$ is the solution of $c(u_0,s_0)=0$ with $0\leq u_0\leq 1$. Numerically, we observe that the leading-twist terms are dominant for the LCSRs of the HFFs, agreeing well with the usual $\delta$-power counting rule. Thus, those HFFs shall provide us with a useful platform in testing the properties of the leading-twist LCDAs via comparisons with the data or predictions in other theoretical approaches.

\section{Numerical Analysis}\label{sec:III}
For the numerical analysis, the input parameters are taken as follows. The mass of mesons are $m_D = 1.865~{\rm GeV}$, $m_\rho = 0.775~{\rm GeV}$, $m_\omega = 0.782~{\rm GeV}$ and $m_{K^*} = 0.892~{\rm GeV}$. The $c$-quark pole mass $m_c=1.28(3)~{\rm GeV}$ is taken from the particle data group~\cite{Tanabashi:2018oca}. For the decay constants, we take $f_D = 0.204(5)$ for $D$-meson, $f_\rho^\| = 0.198(7)$ and $f_\rho^\bot = 0.160(10)$ for $\rho$ meson, $f_\omega^\| = 0.195(3)$ and $f_\omega^\bot = 0.145(10)$ for $\omega$ meson, $f_{K^*}^\| = 0.226(28)$ and $f_{K^*}^\bot = 0.185(10)$ for $K^*$ meson~\cite{Ball:2004rg}. The Cabibbo-Kobayashi-Maskawa matrix elements $|V_{cd}| = 0.216$ and $|V_{cs}| = 0.997$.

\subsection{LCDAs and $D\to V$ HFFs}\label{sec:HFFs}
Within the QCD LCSR framework, HFFs will be expressed by different twist LCDAs due to the same method for handling correlation function to get the TFFs. The resultant HFFs contain twist-2,3,4 LCDAs. Next, we will discuss the associated LCDAs and parameters.

For the leading twist LCDAs, its conformal expansion can be expressed in terms of Gegenbauer polynomials,
\begin{align}
\phi_{2;V}^{\|,\bot}(u,\mu^2) = \phi_{\rm asy} (u) \left[1 + \sum a_n(\mu^2) C_n^{3/2}(\xi)\right].
\end{align}
The $\phi_{\rm asy} (u) = 6u\bar u$ stands for the asymptotic DA. The $\phi_{2;V}^{\|,\bot}(u,\mu^2)$ will equal to $\phi_{\rm asy} (u)$ in the limit $\mu^2\to \propto$. To make a comparison with other theoretical and experimental predictions, the twist-2,3,4 LCDAs' moments and coupling constants can refer to P. Ball~\cite{Ball:1998kk}, which are calculated within SVZ QCD sum rule taken by many theoretical groups. The analytical expression and values are listed in the Appendix.

Then, there are two internal parameters, i.e. continuum threshold $s_0$ and Borel windows $M^2$. The former is a demarcation for the $D$-meson ground state and higher mass contributions. Specifically, we take the continuum thresholds $s_0$ for $D\to V$ HFFs ${\cal D}_{V,0}(q^2)$, ${\cal D}_{V,1}(q^2)$, ${\cal D}_{V,2}(q^2)$ and ${\cal D}_{V,t}(q^2)$ as: $s_{\rho,0}=4.0(3)$, $s_{\rho,1}=4.0(3)$, $s_{\rho,2}=4.0(3)$, $s_{\rho,t}=4.5(3)$, $s_{\omega,0}=3.6(3)$, $s_{\omega,1}=6.5(3)$, $s_{\omega,2}=4.0(3)$, $s_{\omega,t}=4.0(3)$, $s_{K^*,0}=4.0(3)$, $s_{K^*,1}=6.0(3)$, $s_{K^*,2}=4.0(3)$ and $s_{K^*,t}=3.7(3)$.

To determine the Borel parameters for the $D\to V$ HFFs, we adopt the following three criteria:
\begin{itemize}
\item We require the continuum contribution to be less than 35\% of the total LCSR.
\item We require all the high-twist LCDAs' contributions to be less than 15\% of the total LCSR.
\item The derivatives of LCSRs for HFFs with respect to $(-1/M^2)$ give four LCSRs for the $D$-meson mass $m_D$. We require the predicted $D$-meson mass to be fulfilled in comparing with the experiment one, i.e. $|m_D^{\rm th} - m_D^{\rm exp} |/ m_D^{\rm exp}\leq 0.1\%$.
\end{itemize}
Thus, the obtained Borel windows $M^2 (\rm{GeV}^2)$ are: $M^2_{\rho,0}=2.5(3)$, $M^2_{\rho,1}=4.0(3)$, $M^2_{\rho,2}=3.5(3)$, $M^2_{\rho,t}=3.0(3)$, $M^2_{\omega,0}=2.5(3)$, $M^2_{\omega,1}=4.8(3)$, $M^2_{\omega,2}=3.0(3)$, $M^2_{\omega,t}=3.0(3)$, $M^2_{K^*,0}=2.5(3)$, $M^2_{K^*,1}=6.0(3)$, $M^2_{K^*,2}=4.0(3)$ and $M^2_{K^*,t}=2.7(3)$.

\begin{table}[t]
\centering
\small
\caption{The fitted parameters $a_{0;1;2}^V$ for the HFFs $\mathcal{D}_{V,\sigma}$, where all input parameters are set to be their central values.}\label{tab:analytic}
\renewcommand{\arraystretch}{1.5}
\addtolength{\arraycolsep}{10pt}
\begin{tabular}{c c c c c c}
\hline
 &${\cal D}_{V,0}$ & ${\cal D}_{V,1}$ & ${\cal D}_{V,2}$ & ${\cal D}_{V,t}$
\\ \hline
$a_0^\rho$    & 1.841    & 1.187    & 4.257    & $0.913$
\\
$a_1^\rho$    & $-68.95$ & $-5.177$ & $-137.2$ & $-18.49$
\\
$a_2^\rho$    & $-879.8$ & $-88.14$ & 1774     & $160.2$
\\
$\Delta_\rho$ & 0.000    & 0.008    & 0.042    & $0.000$
\\
$a_0^\omega$   & 1.786    & 0.763     &  4.666   & 0.868
\\
$a_1^\omega$   & $-68.41$ & $-1.125$  & $-162.6$ & $-18.46$
\\
$a_2^\omega$   & 883.8    & $-22.53$  & 2163     & 170.1
\\
$\Delta_\omega$& 0.000    & 0.001     & 0.050    & 0.000
\\
$a_0^{K^*}$    & 1.937    & $0.941$   & 5.074   & $0.975$
\\
$a_1^{K^*}$    & $-91.04$ & $2.976$   & $-209.4$  & $-21.97$
\\
$a_2^{K^*}$    &  1438    & $-70.30$  & $3545$    & $181.0$
\\
$\Delta_{K^*}$ & 0.000    & $0.001$   & $0.031$ & $0.000$
\\
\hline
\end{tabular}
\addtolength{\arraycolsep}{-10pt}
\end{table}

\begin{figure*}[t]
\begin{center}
\includegraphics[width=0.33\textwidth]{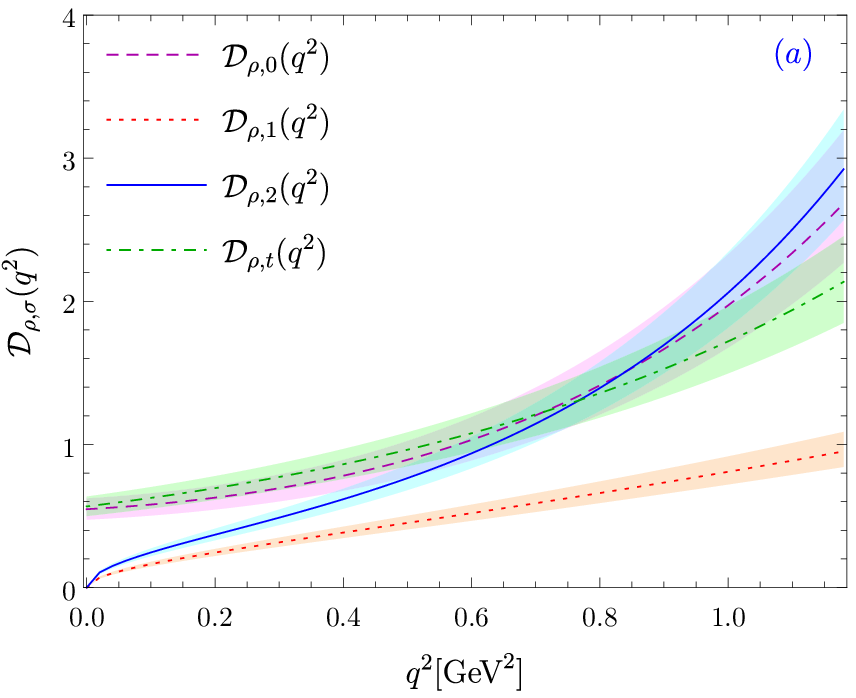}\includegraphics[width=0.33\textwidth]{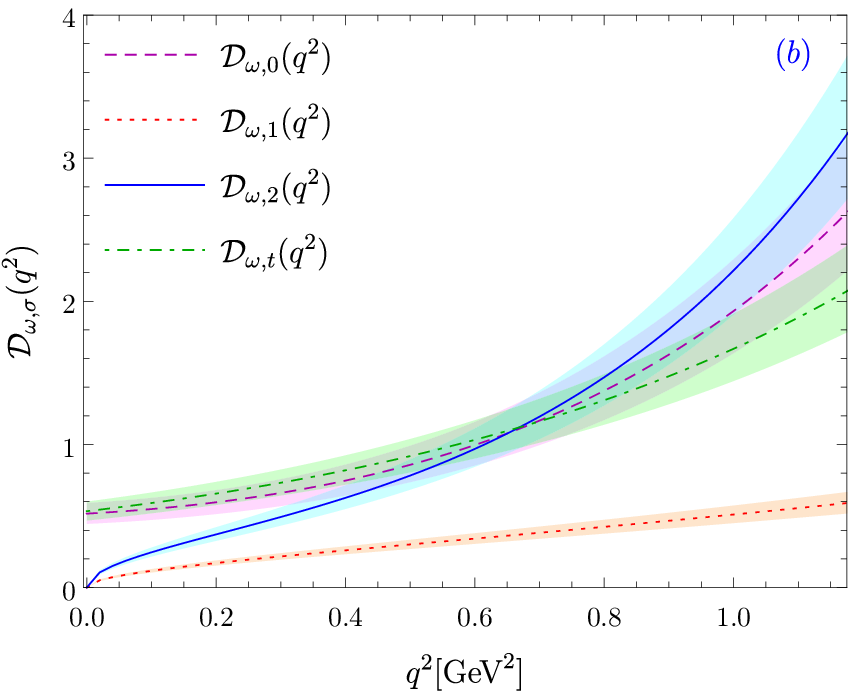}\includegraphics[width=0.33\textwidth]{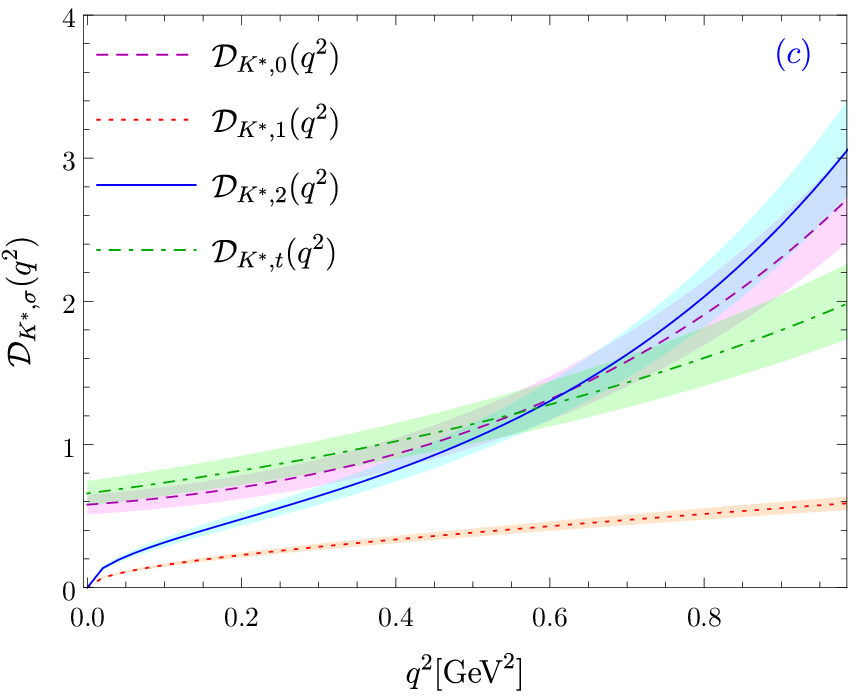}
\end{center}
\caption{The extrapolated LCSR predictions HFFs ${\cal D}_{V,\sigma}(q^2)$ for the $D \to V$ with $V=\rho, \omega, K^*$-mesons. The solid lines represent the center values and the shaded bands corresponds to their uncertainties. The maximum extrapolated physically allowable point $q^2$ are $q^2_{\rho,{\rm max}} = (m_D-m_\rho)^2 \simeq 1.18~\rm{GeV}^2$, $q^2_{\omega,{\rm max}} = (m_D-m_\omega)^2 \simeq 1.17~\rm{GeV}^2$ and $q^2_{K^*,{\rm max}} = (m_D-m_K^*)^2 \simeq 0.98~\rm{GeV}^2$ for $\rho$, $\omega$ and $K^*$-mesons, respectively.}
\label{HFF:P0t}
\end{figure*}

The reliable regions for the $D$-meson semileptonic decays within LCSR approach can be set at $0\leq q^2\leq q^2_{\rm LCSR, max}\approx 0.8~\rm{GeV}^2$. Meanwhile, the allowable physical range of the momentum transfer is $0\leq q^2\leq q^2_{V,{\rm max}}$ with
\begin{align}
q^2_{\rho, {\rm max}}  &= (m_D-m_\rho)^2   \simeq 1.18~\rm{GeV}^2, \nonumber\\
q^2_{\omega,{\rm max}} &= (m_D-m_\omega)^2 \simeq 1.17~\rm{GeV}^2, \nonumber\\
q^2_{K^*,{\rm max}}    &= (m_D-m_{K^*})^2  \simeq 0.98~\rm{GeV}^2, \nonumber
\end{align}
for $\rho$, $\omega$ and $K^*$-mesons, respectively. Then, we use the SSE to do the extrapolation for the HFFs based on the analyticity and unitarity consideration. The extrapolation of the HFFs satisfies the following parameterized formulas,
\begin{eqnarray}
\mathcal{D}_{V,0}(t) &=&\frac{1}{B(t) \sqrt{z(t,t_-)} \phi_T^{V-A}(t)} \sum_{k=0,1,2} a_k^{V,0} z^k ,  \\
\mathcal{D}_{V,1}(t) &=&\frac{\sqrt{-z(t,0)}}{B(t) \phi_T^{V-A}(t)} \sum_{k=0,1,2} a_k^{V,1} z^k ,   \\
\mathcal{D}_{V,2}(t) &=&\frac{\sqrt{-z(t,0)}}{B(t) \sqrt{z(t,t_-)} \phi_T^{V-A}(t)} \sum_{k=0,1,2} a_k^{V,2} z^k ,\\
\mathcal{D}_{V,t}(t) &=&\frac{1}{B(t)\phi_L^{V-A}(t)} \sum_{k=0,1,2} a_k^{V,t} z^k ,
\end{eqnarray}
where $\phi_I^X(t)=1$, $\sqrt{-z(t,0)}=\sqrt{q^2}/m_D$, $B(t)=1- q^2/m_{\sigma}^2$, $\sqrt{z(t,t_-)}=\sqrt{\lambda}/m_D^2$, and
\begin{eqnarray}
z(t)=\frac{\sqrt{t_+ - t}-\sqrt{t_+ - t_0}}{\sqrt{t_+ - t}+\sqrt{t_+ - t_0}}\nonumber
\end{eqnarray}
with $t_\pm=(m_D \pm m_V)^2$ and $t_0=t_+(1-\sqrt{1-t_-/t_+})$.

The parameters $a_k^{\sigma}$ can be determined by requiring the ``quality'' of fit ($\Delta_V$) to be less than one, which is defined as
\begin{equation}
\Delta_V=\frac{\sum_t\left|\mathcal{D}_{V,\sigma}(t)-\mathcal{D}_{V,\sigma}^{\rm fit}(t)\right|} {\sum_t\left|\mathcal{D}_{V,\sigma}(t)\right|}\times 100, \label{delta}
\end{equation}
where $t\in[0,0.02,\cdots,0.58,0.8]~{\rm GeV}^2$. We put the determined parameters $a_{k}^{V,\sigma}$ in Table~\ref{tab:analytic}, in which all the input parameters are set to be their central values.

\begin{figure*}[t]
\begin{center}
\includegraphics[width=0.33\textwidth]{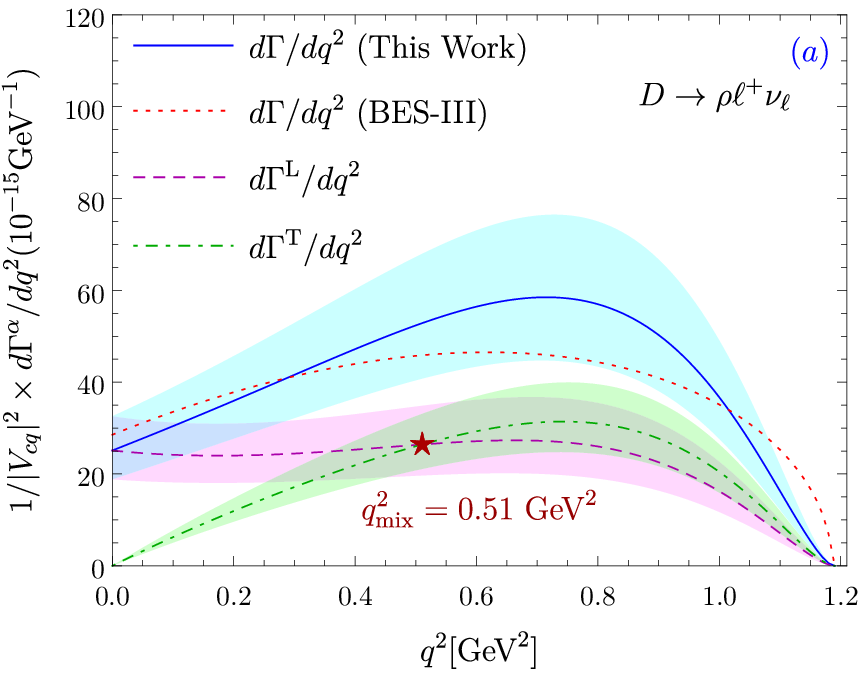}\includegraphics[width=0.33\textwidth]{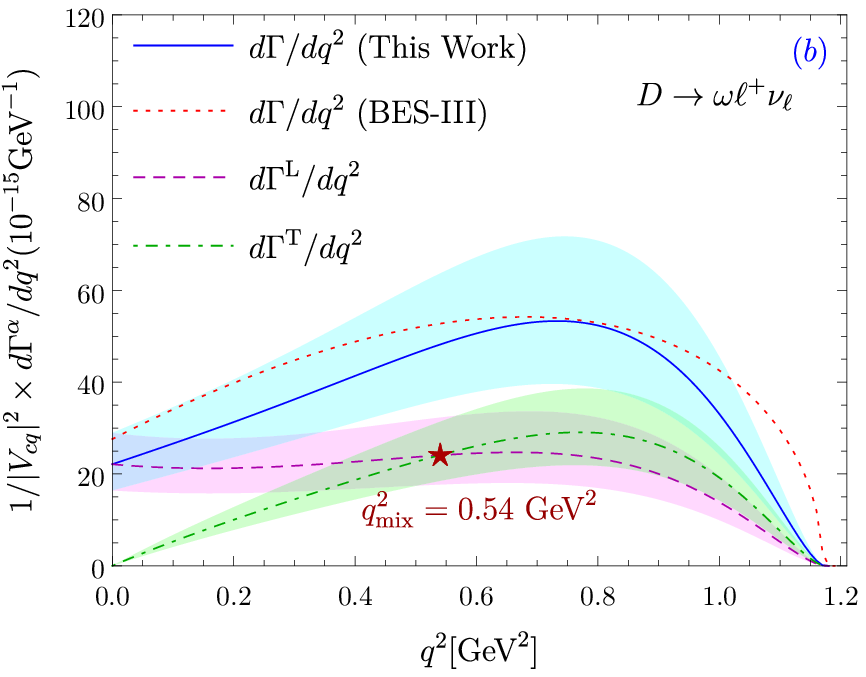}\includegraphics[width=0.33\textwidth]{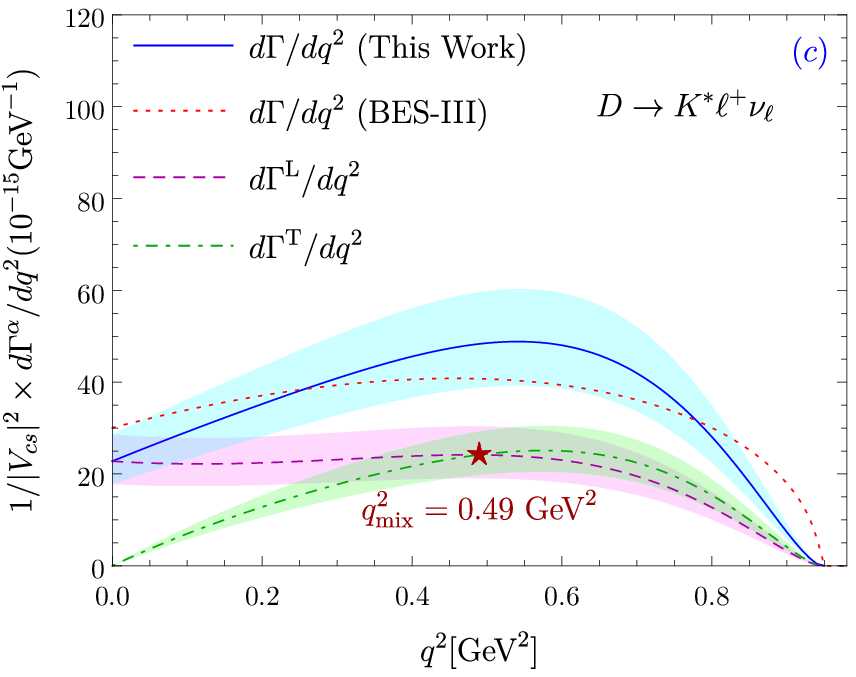}
\end{center}
\caption{The LCSR predictions for the polarized differential decay widths $1/|V_{cq}|^{2} \times d \Gamma^{\rm L,T}/{d q^2}$ and the total differential decay widths $1/|V_{cq}|^{2} \times d \Gamma/{d q^2}$ for $\rho, \omega, K^*$-mesons, in which the dashed, dotted and solid line represent the corresponding central values, and the shaded band are the squared average of all the input parameters. For comparison, we also present the BES-III predictions~\cite{Ablikim:2018qzz,Ablikim:2015gyp,Ablikim:2018lmn}.}
\label{fig:dGLT}
\end{figure*}

The extrapolated HFFs in whole $q^2$-region are presented in Fig.~\ref{HFF:P0t}, where the shaded bands are uncertainties from various input parameters. The shape of HFFs for the three vector mesons are similar due to the same analytic expression and a little varied input parameters. We can see $\mathcal{D}_{V,(1;2)}=0$ at $q^2=0~{\rm GeV}^2$, which are caused by the coefficient $q^2$ of $\mathcal{D}_{V,(1;2)}(q^2)$. The $q^2$ coefficient also depresses the error of HFFs $\mathcal{D}_{V,(1;2)}(q^2)$ for the smaller $q^2$, which can be directly seen from Fig.~\ref{HFF:P0t}. Meanwhile, this depressed effect can be directly transmitted to the differential transversal decay width in Fig.~\ref{fig:dGLT} seen in the next subsection.

\subsection{$D$-meson Semileptonic Decays}\label{sec:App}
The HFFs extracted from the LCSRs are employed to study the $D$-meson semileptonic decay, i.e. the decay width, branching fractions, longitudinal and transverse polarization, forward-backward asymmetry and lepton-side convexity parameter which are frequently used for precision test of the SM and search for new physics beyond SM.

\begin{table}[t]
\centering
\small
\caption{The total decay widths $\Gamma/|V_{cq}|^{2} $, $ \Gamma^{\rm L}/|V_{cq}|^{2}$ and $\Gamma^{\rm T}/|V_{cq}|^{2} $ (in the unit $ 10^{-15}~{\rm{GeV}}$) for the central values.}
\renewcommand{\arraystretch}{1.5}
\begin{tabular}{lccccl}
\\[-2ex]
\hline
                                & $\Gamma/|V_{cq}|^{2} $ & $ \Gamma^{\rm L}/|V_{cq}|^{2}$ & $\Gamma^{\rm T}/|V_{cq}|^{2} $              \\ \hline
$D\to\rho \ell^+\nu_\ell$       & $49.564^{+14.814}_{-11.633}$           & $26.299^{+8.682}_{-6.801}$              & $23.265^{+6.131}_{-4.832}$ \\
$D\to\omega  \ell^+\nu_\ell$    & $44.108^{+14.610}_{-11.187}$           & $23.320^{+8.059}_{-6.202}$              & $20.788^{+6.551}_{-4.986}$ \\
$D\to K^* \ell^+\nu_\ell$       & $33.631^{+7.940}_{-6.700}$             & $18.539^{+4.763}_{-3.941}$              & $15.092^{+3.177}_{-2.759}$ \\
\hline
\end{tabular}
\label{tab:GALL}
\end{table}

\subsubsection{Decay width}

%%%%%%%%%%%%%%%%%%%%%%%%%%%%%%%%%
\begin{table*}[t]
\centering
\caption{Ratio $\Gamma^{\rm L}/\Gamma^{\rm T}$ for the $D \to V(\rho,\omega,K^*) \ell^+\nu_\ell$ semileptonic decays, where the uncertainties are the squared average of all the input parameters. The theoretical and lattice results are listed as a comparison. Note that the lepton mass is ignored in HM$\chi$T.~\cite{Fajfer:2005ug}. For convenience, we list it in the $D \to Xe^+\nu_e$ case due to the electron mass is too small to be ignored.}
\renewcommand{\arraystretch}{1.5}
\addtolength{\arraycolsep}{15pt}
\small
\begin{tabular}{lccccccc}
\\[-2ex]
\hline
                         & This work               & HM$\chi$T\cite{Fajfer:2005ug}& CCQM~\cite{Ivanov:2019nqd} & CQM~\cite{Melikhov:2000yu} & LCSR\cite{Wang:2002zba} &QCDSR\cite{Ball:1991bs} & LQCD\cite{Allton:1994ui} \\ \hline
$D\to\rho e^+\nu_e$      & $1.130^{+0.095}_{-0.133}$ & 1.10                       & 1.13                       & 1.16                    & 1.17(9)                    &0.86(6)& -                         \\
$D\to\rho\mu^+\nu_\mu$   & $1.119^{+0.095}_{-0.132}$ & -                      & 1.04                       & -                          & -                       & -                        & -                                         \\
$D\to\omega e^+\nu_e$    & $1.122^{+0.042}_{-0.075}$ & 1.10                       & 1.10                       & -                          & -                        & -                        & -                                         \\
$D\to\omega\mu^+\nu_\mu$ & $1.110^{+0.042}_{-0.074}$ & -                      & 1.02                       & -                          & -                        & -                        & -                                         \\
$D\to K^* e^+\nu_e$      & $1.228^{+0.061}_{-0.074}$ & 1.13                       & 1.18                       &1.28                        &1.15(10)                  & -
&1.2(3)                \\
$D\to K^*\mu^+\nu_\mu$   & $1.212^{+0.060}_{-0.073}$ & -                      & 1.07                       &-                            &-                          & -                        &-\\
\hline
\end{tabular}
\addtolength{\arraycolsep}{-15pt}
\label{tab:GLT}
\end{table*}

In this part, we probe the decay width of $D \to V$ semilepton decay by applying the Eqs.~\eqref{difftot}, \eqref{diffL} and \eqref{diffT}. Firstly, we present the LCSR predictions for the polarization differential decay widths $1/|V_{cq}|^{2} \times d \Gamma^{\rm L}/{d q^2}$, $1/|V_{cq}|^{2} \times d \Gamma^{\rm T}/{d q^2}$ and the total differential decay widths $1/|V_{cq}|^{2} \times d \Gamma/{d q^2}$ in Fig.~\ref{fig:dGLT}, in which the dashed, dotted and solid line represent the corresponding central values, the uncertainties are a result of the squared average of all input parameters.

For the central lines of the differential decay width in Fig.~\ref{fig:dGLT}, we find that there is a similar behavior for all of the differential decay width with $|V_{cq}|$ independent of $D \to V\ell^+\nu_\ell$ semileptonic decays. Both the total differential width and transversal differential width increase first and then decrease with $q^2$. The longitudinal differential width is almost unchanged from the small to middle $q^2$-region, while it drops in the large $q^2$-region sharply. Besides, the longitudinal differential width dominate the small $q^2$-region, while the transversal differential width dominate the large $q^2$-region. The position of the alternating point of the dominant $q^2$-region is near the midpoint of the whole physical feasible region, which are represented by the red stars in Fig.~\ref{fig:dGLT}, i.e. $q^2_{{\rm mix},(\rho,\omega, K^*)}= (0.51,0.54,0.49)~{\rm GeV}^2$.

The three figures imply that the total width decreases as the final meson mass increases, which is intuitive from Table~\ref{tab:GALL}. There are three main reasons to justify this:
\begin{itemize}
\item[(i)]The physical feasible regions ($q^2$) decreases from the left to right in Fig.~\ref{fig:dGLT}, which is caused by the increasing mass of final meson;
\item[(ii)]The peak of the longitudinal and transversal differential width decreases from the left to right panel of Fig.~\ref{fig:dGLT} ;
\item[(ii)]The trend of curves for longitudinal and transverse differential width are almost the same for the three channels in Fig.~\ref{fig:dGLT}.
\end{itemize}

\begin{table*}[tb]
\centering
%\footnotesize
\renewcommand{\arraystretch}{1.5}
\addtolength{\arraycolsep}{20pt}
\caption{Branching fractions for semileptonic $D$ decays i.e. $D \to V(\rho,\omega,K^*) \ell^+ \nu_\ell$ (in unit: $10^{-3}$ for $\rho$ and $\omega$-mesons; $10^{-2}$ for $K^*$ meson), where the uncertainties are the squared averages of all the input parameters. In which the current theoretical and experimental results in references are also listed as a comparison. Note that the lepton mass is ignored in HM$\chi$T.~\cite{Fajfer:2005ug}. For convenience, we list it in the $D \to Xe^+\nu_e$ case due to the electron mass is too small to be ignored.}\label{tab:BrD2V}
\begin{tabular}{l ccccc}\\[-2ex]\hline
                                                                & $D^0\to\rho^- e^+ \nu_e$        & $D^0\to\rho^-\mu^+\nu_\mu$      & $D^+\to\rho^0 e^+ \nu_e$           & $D^+\to\rho^0\mu^+\nu_\mu$      & $D^+\to\omega e^+ \nu_e$      \\ \hline
This Work                                                       & $1.440^{+0.277}_{-0.250}$ & $1.432^{+0.274}_{-0.248}$ & $1.827^{+0.351}_{-0.317}$    & $1.816^{+0.348}_{-0.314}$ & $1.740^{+0.482}_{-0.399}$   \\
HM$\chi$T\cite{Fajfer:2005ug}                                   & 2.0                      & -                      & 2.5                         & -                      & 2.5                        \\
CCQM\cite{Ivanov:2019nqd}                                       & 1.62                      & 1.55                      & 2.09                         & 2.01                      & 1.85                        \\
LFQM\cite{Cheng:2017pcq}                                        & -                         & -                         & -                            & -                         & $2.1(2)$                    \\
$\chi$UA~\cite{Sekihara:2015iha}                                & 1.97                      & 1.84                      & 2.54                         & 2.37                      & 2.46                        \\
LCSR\cite{Wu:2006rd}                                            & $1.81^{+0.18}_{-0.13}$    & $1.73^{+0.17}_{-0.13}$    & $2.29^{+0.23}_{-0.16}$       & $2.20^{+0.21}_{-0.16}$    & $1.93^{+0.20}_{-0.14}$      \\
LQCD\cite{Allton:1994ui}                                   & -                         & -                         & 2.23(70)                     & 2.13(64)                  & -                           \\
BES-III\cite{Ablikim:2018qzz,Ablikim:2015gyp,Ablikim:2018lmn}    & $1.445(70)$               & -                         & $1.860(93)$                  & -                         & $1.63(14)$                  \\
CLEO\cite{CLEO:2011ab,Coan:2005iu,Briere:2010zc}                & $1.77(16)$                & -                         & $2.17(12)(^{+0.12}_{-0.22})$ & -                         & $1.82(19)$                  \\
PDG\cite{Tanabashi:2018oca}                                     & -                         & -                         & -                            & $2.4(4)$                  & -                           \\ \hline
                                                               & $D^+\to\omega\mu^+\nu_\mu$ & $D^0\to K^{*-}e^+\nu_e$          & $D^0\to K^{*-}\mu^+\nu_\mu$      & $D^+\to \bar K^{*0} e^+\nu_e$        & $D^+\to\bar{K}^{*0}\mu^+\nu_\mu$ \\ \hline
This Work                                                      & $1.728^{+0.479}_{-0.397}$  & $2.082^{+0.334}_{-0.314}$ & $2.066^{+0.330}_{-0.310}$ & $5.282^{+0.847}_{-0.796}$     & $5.242^{+0.838}_{-0.787}$  \\
HM$\chi$T\cite{Fajfer:2005ug}                                  & -                       & 2.2                      & -                      & 5.6                          & -                       \\
CCQM\cite{Ivanov:2019nqd}                                      & 1.78                       & 2.96                      & 2.80                      & 7.61                          & 7.21                       \\
LFQM\cite{Cheng:2017pcq}                                       & $2.0(2)$                   & -                         & -                         & $7.5(7)$                      & $7.0(7)$                   \\
$\chi$UA~\cite{Sekihara:2015iha}                               & 2.29                       & 2.15                      & 1.98                      & 5.56                          & 5.12                       \\
LCSR\cite{Wu:2006rd}                                           & $1.85^{+0.19}_{-0.13}$     & $2.12(9)$    & $2.01^{+0.09}_{-0.08}$    & $5.37^{+0.24}_{-0.23}$        & $5.10^{+0.23}_{-0.21}$     \\
LQCD\cite{Allton:1994ui}                                  & -                          & -                         & -                         & 6.26(184)                     & 5.95(167)                  \\
BES-III\cite{Ablikim:2018qzz,Ablikim:2015gyp,Ablikim:2018lmn}   & 1.77(29)                         & $2.033(66)$               & -                         & -                             & -                          \\
CLEO\cite{CLEO:2011ab,Coan:2005iu,Briere:2010zc}               & -                          & $2.16(17)$                & -                         & -                             & $5.27\pm 0.16$             \\
PDG\cite{Tanabashi:2018oca}                                    & -                          & -                         & -                         & $5.4(1)$                      & -                          \\ \hline
\end{tabular}
\addtolength{\arraycolsep}{-20pt}
\end{table*}
\begin{figure*}[t]
\begin{center}
\includegraphics[width=0.33\textwidth]{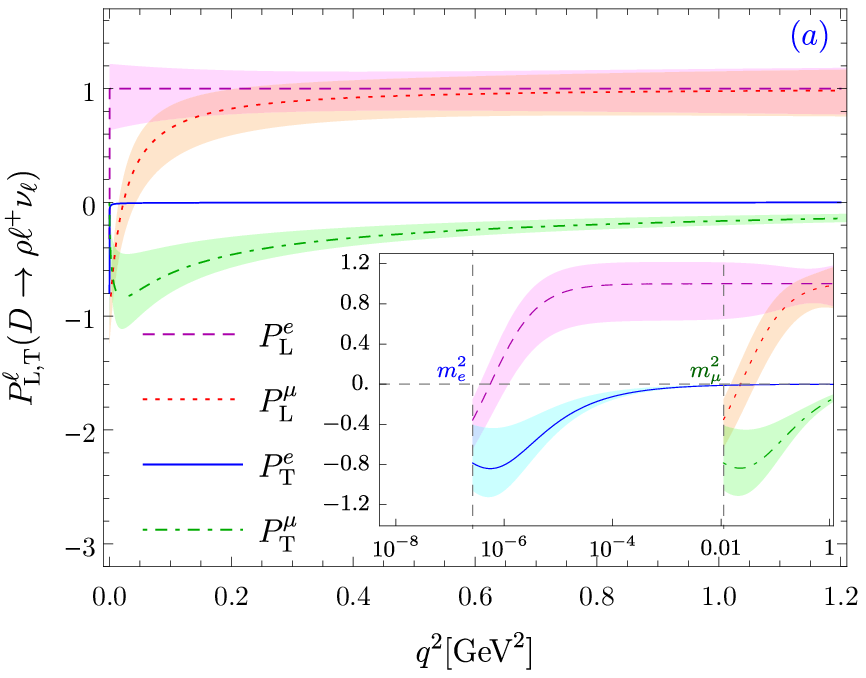}\includegraphics[width=0.33\textwidth]{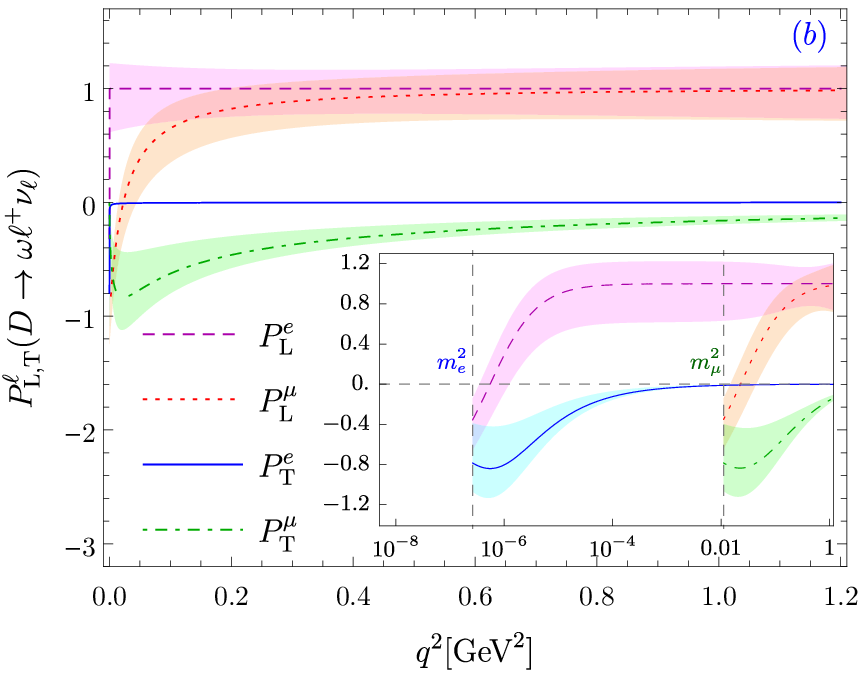}\includegraphics[width=0.33\textwidth]{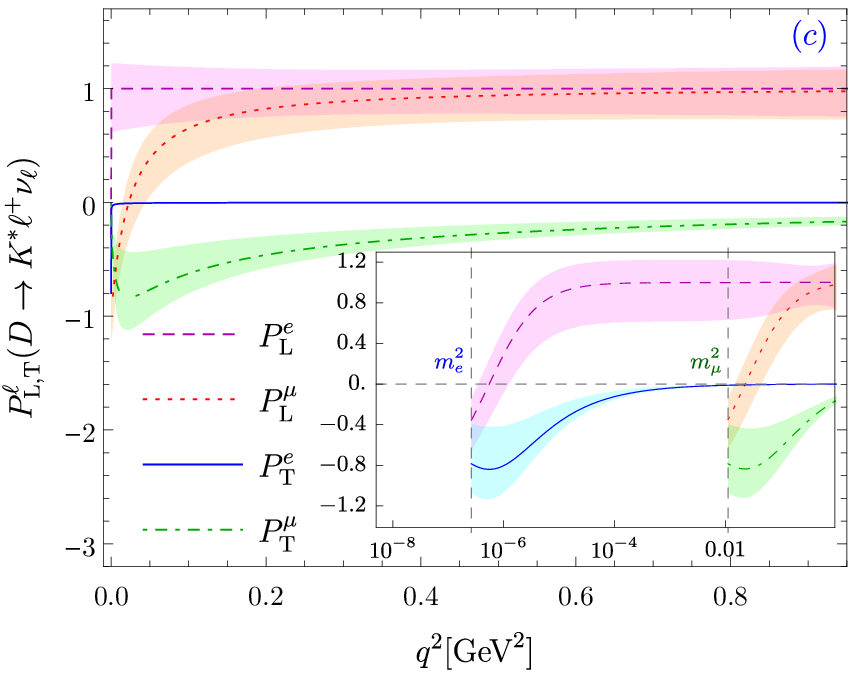}
\includegraphics[width=0.33\textwidth]{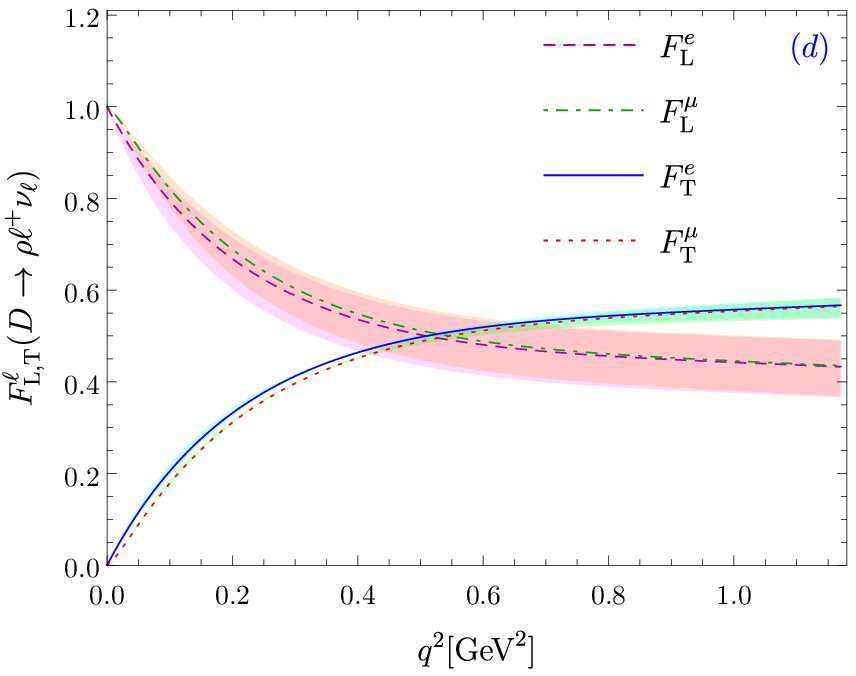}\includegraphics[width=0.33\textwidth]{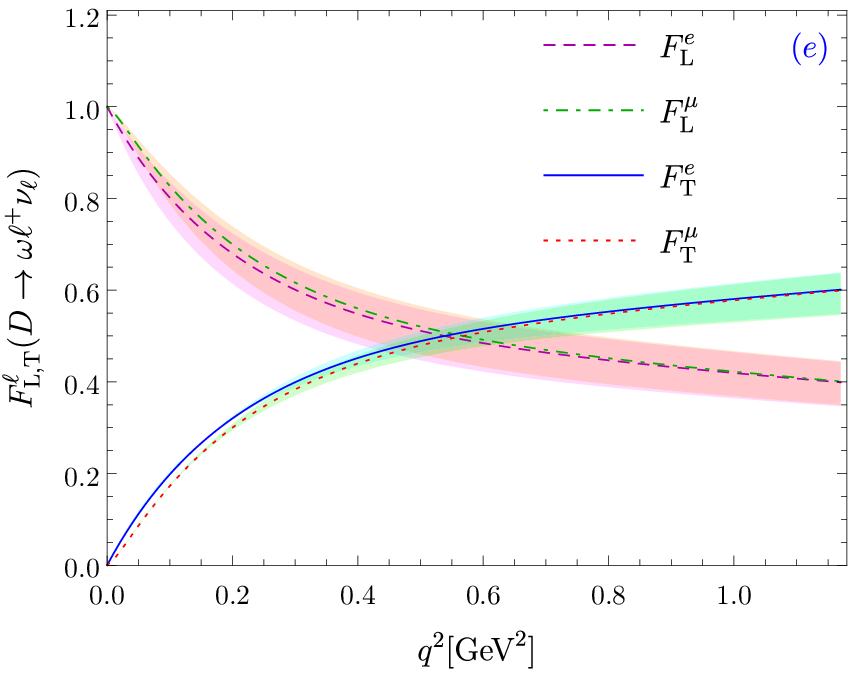}\includegraphics[width=0.33\textwidth]{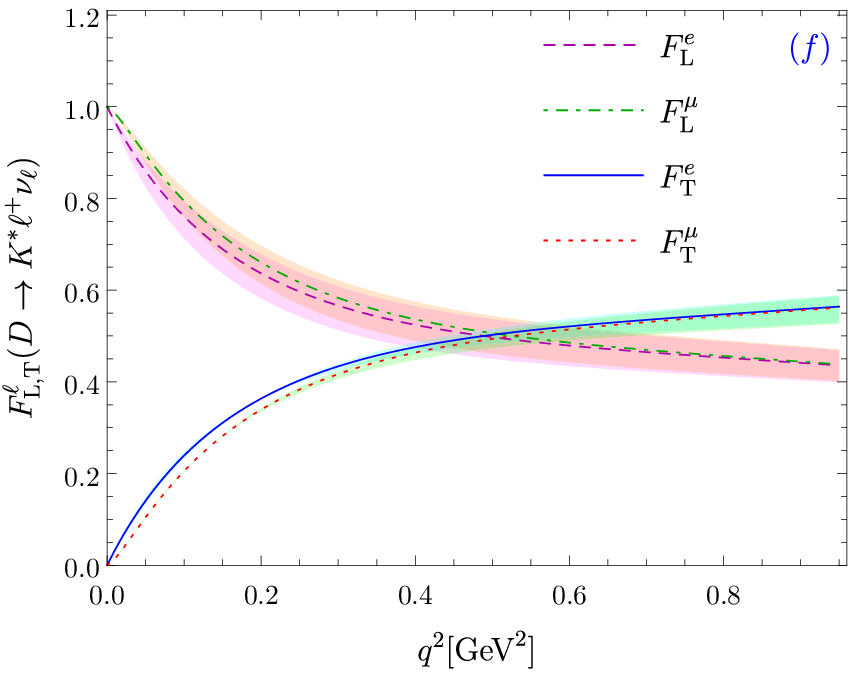}
\end{center}
\caption{The final state polarization $P_{\rm L,T}^\ell$ and $F_{\rm L,T}^\ell$ as a function of $q^2$ for the $D \to V \ell^+\nu_\ell$. Here $P$ and $F$ represent charged lepton and vector meson in the final state, which corresponds to the upper and bottom parts respectively; The symbols $T$ and $L$ stand for longitudinal and transverse fractions; $V$ stands for the $\rho, \omega, K^*$-mesons corresponding to left, medial and right part respectively. In which the dashed-, dotted-, dot-dashed- and solid-line represent the corresponding central values, and the shaded band is the corresponding errors from HFFs.}
\label{Fig:PLTFLT}
\end{figure*}

For comparison, the central values of the total differential width of BES-III~\cite{Ablikim:2018qzz,Ablikim:2015gyp,Ablikim:2018lmn} are also shown in the Fig.~\ref{fig:dGLT}. We observe that the curves of BES-III are in agreement with our predictions in errors. But there is a significantly different for the shape of the center curves, especially for the large $q^2$-region. The main reason is that BES-III use the unipolar point continuation extrapolation method, while HFFs need to use the SSE extrapolation method.

Then we show the total decay widths $\Gamma/|V_{cq}|^{2}$, $\Gamma^{\rm L}/|V_{cq}|^{2}$ and $\Gamma^{\rm T}/|V_{cq}|^{2}$ in the Table~\ref{tab:GALL}. The three kinds of total decay widths decrease with the final meson mass increasing, which is consistent with Fig.~\ref{fig:dGLT}. There is also an interesting phenomenon: it is almost identical for both total decay widths $ \Gamma^{\rm L}/|V_{cq}|^{2}$ and $\Gamma^{\rm T}/|V_{cq}|^{2} $ gaps between different decay channels. We list the ratio $\Gamma^{\rm L}/\Gamma^{\rm T}$ for the $D\to V \ell^+ \nu_\ell$ semileptonic decays in the Table~\ref{tab:GLT}. As a comparison, we also present other theoretical predictions, i.e. Heavy Meson and Chiral Lagrangians (HM$\chi$T)~\cite{Fajfer:2005ug}, Covariant Confining Quark Model (CCQM)~\cite{Ivanov:2019nqd}, Covariant Quark Model (CQM)~\cite{Melikhov:2000yu}, LCSR~\cite{Wang:2002zba}, QCD sum rule(QCDSR)~\cite{Ball:1991bs} and Lattice QCD (LQCD)~\cite{Allton:1994ui}. All of our predictions for the ratio $\Gamma^{\rm L}/\Gamma^{\rm T}$ agree with that of the CCQM within errors. Although the rest of the theoretical predictions are incomplete for this ratio $\Gamma^{\rm L}/\Gamma^{\rm T}$, again, our results are in good agreement with them within errors, except for QCDSR results.

As a further step, we calculate the branching fractions of $D\to V \ell^+ \nu_\ell$ by employing $\tau({D^0}) = 0.410(2)~{\rm ps}$ and $\tau({D^+}) = 1.040(7)~{\rm ps}$, the results are collected in Table~\ref{tab:BrD2V}. Compared with other theoretical and LQCD~\cite{Allton:1994ui} predictions, our results are small, which is more consistent with the BES-III~\cite{Ablikim:2018qzz,Ablikim:2015gyp,Ablikim:2018lmn} experiment within errors.
The reasons are that we adopt HFFs to deal with the $D \to V$ hadronic matrix elements, where HFFs are calculated by the QCD LCSR approach and the corresponding physical observations are further investigated. Compared with the traditional TFFs parameterized method for the hadronic matrix elements, HFFs parameterized method has some advantages, such as:
\begin{itemize}
\item As mentioned in the introduction of our paper: ``HFFs decompose it by applying the off-shell $W$-boson polarization vectors, which brings a good polarization property, i.e. researching on tracking polarization''.
\item  According to the diagonalizable unitarity relations, one can get the dispersive bound for the HFF parametrization.
\end{itemize}
Besides, there are many theoretical approaches in dealing with the FFs for the $D \to V$ decays processes, such as LCSR used in this paper, the lattice QCD (LQCD), and the perturbative QCD (pQCD), and so on. The pQCD, LCSR, and LQCD approachs are valid in the lower $q^2$-region, in the lower and intermediate $q^2$-region, and in the higher $q^2$-region, respectively. The LCSR approach has the advantage that it can be extrapolated to the whole $q^2$-region, and be provided as an important bridge for connecting various approaches.

\begin{figure*}[t]
\begin{center}
\includegraphics[width=0.33\textwidth]{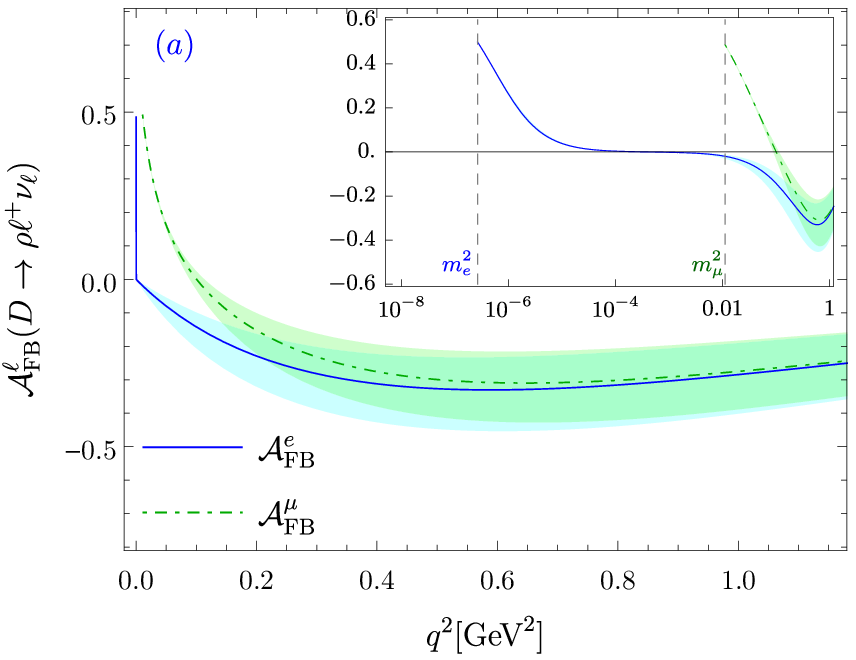}\includegraphics[width=0.33\textwidth]{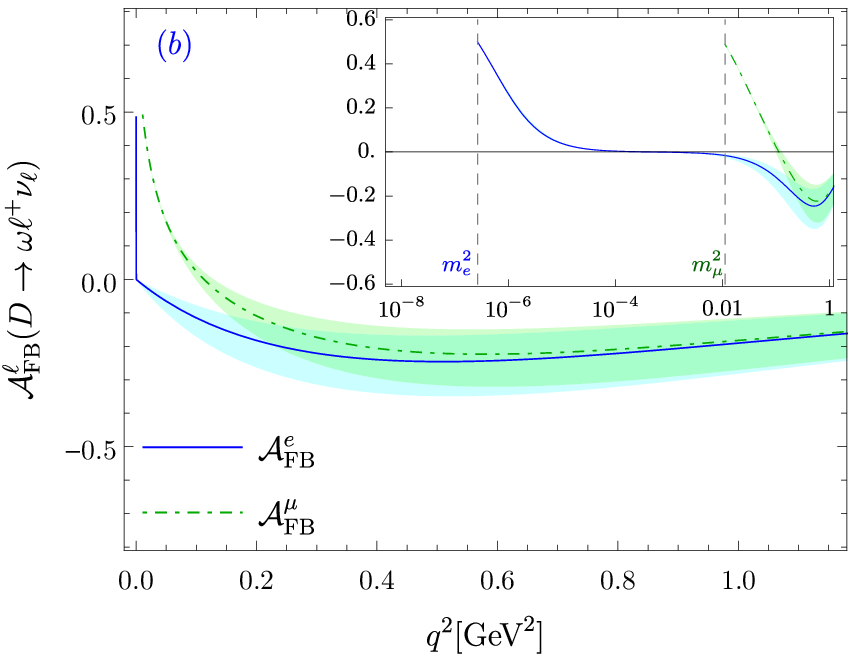}\includegraphics[width=0.33\textwidth]{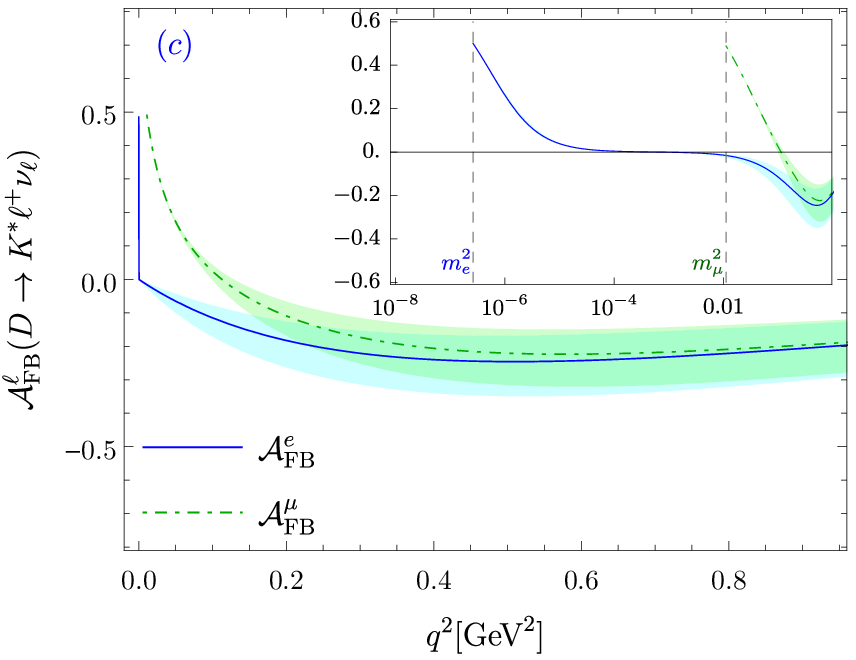}
\includegraphics[width=0.33\textwidth]{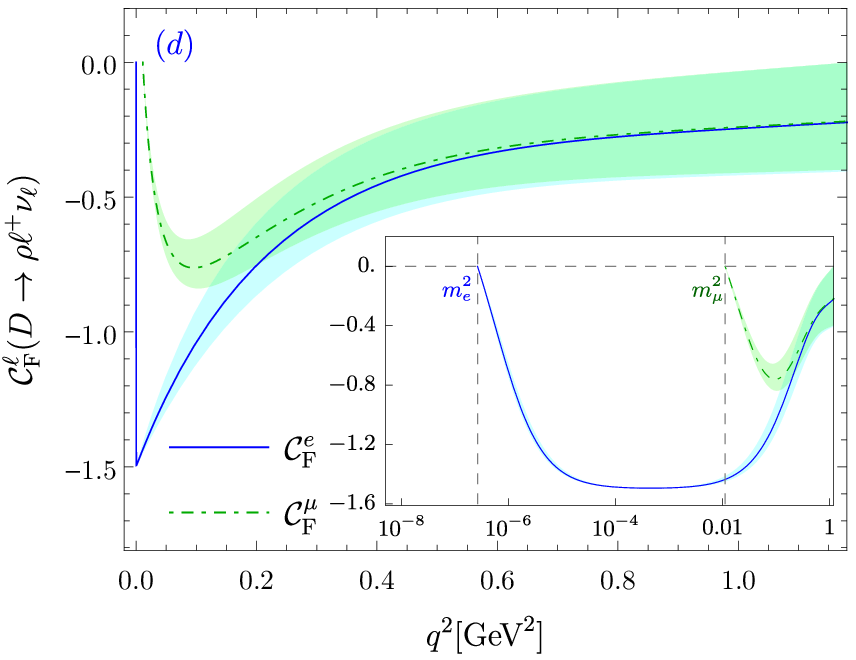}\includegraphics[width=0.33\textwidth]{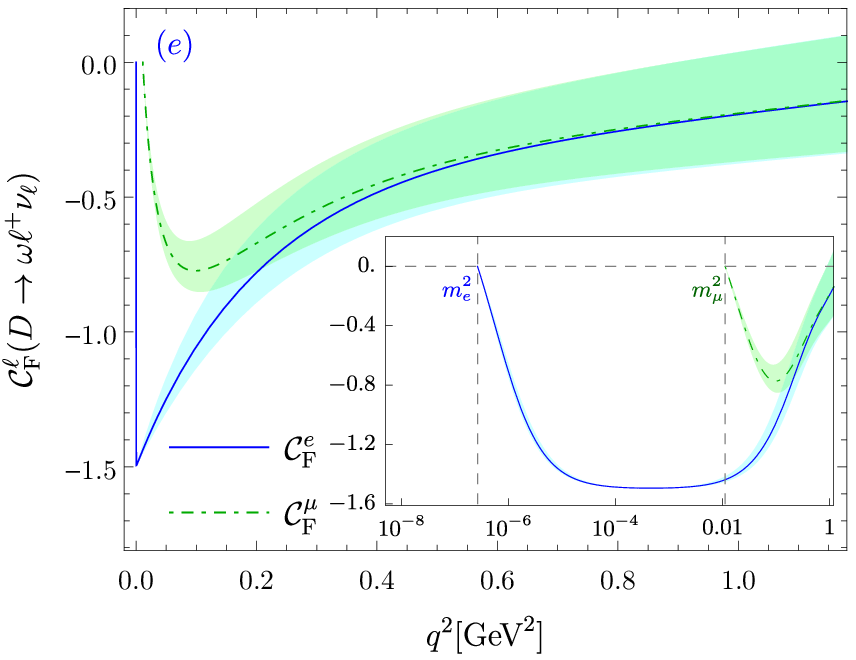}\includegraphics[width=0.33\textwidth]{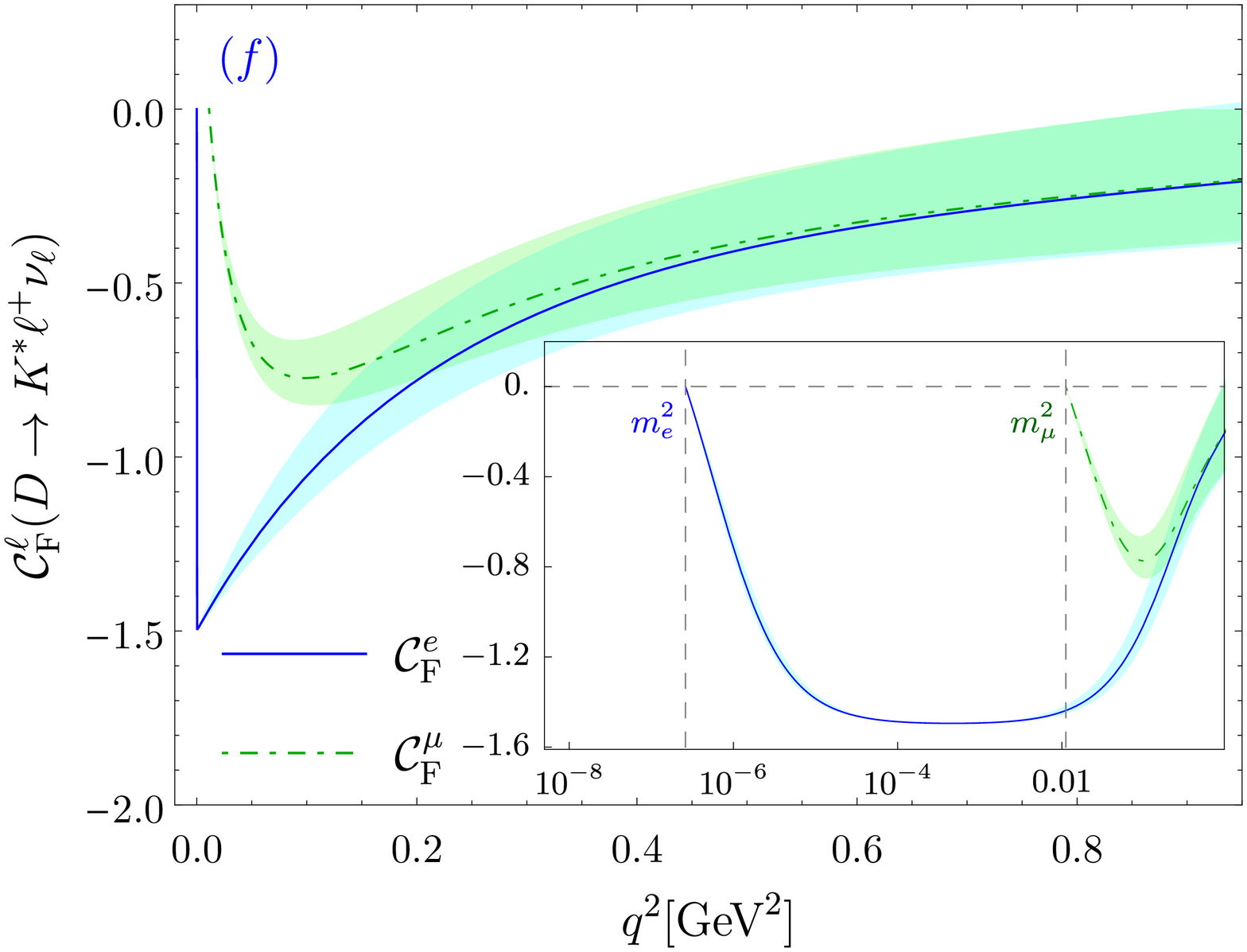}
\end{center}
\caption{Forward-backward asymmetry ${\cal A}_{\rm FB}^\ell(q^2)$ and the lepton-side ${\cal C}_{\rm F}^\ell(q^2)$ convexity parameter as a function of $q^2$ for the $D \to V(\rho,\omega,K^*)\ell^+\nu_\ell$. The lines are their central values and the shaded bands are their errors. The meaning of corresponding representations can refer to Fig.~\ref{Fig:PLTFLT}.}
\label{Fig:AFBCFl}
\end{figure*}

\subsubsection{Polarization observations}

Due to the current experimental conditions, it is difficult to measure the $q^2$ dependence of polarization observation. However, it is very important to study the $q^2$ dependence on these observable. On the one hand, it can facilitate the comparison among different theories; on the other hand, it also provides references for experimental research on $q^2$ dependence and more details for exploring new physics.

We firstly show final state polarization $P_{\rm L,T}^\ell$ and $F_{\rm L,T}^\ell$ in Fig.~\ref{Fig:PLTFLT}.

\begin{table*}[t]
\centering
\small
\caption{The mean values for longitudinal and transverse polarizations fraction of final lepton and vector mesons, forward-backward asymmetry and the lepton-side convexity parameter for positron and muon modes, where the uncertainties are the squared average of all the input parameters.}
\renewcommand{\arraystretch}{1.5}
\addtolength{\arraycolsep}{20pt}
\begin{tabular}{lcccccc}
\\[-2ex]
\hline
                                   & \multicolumn{2}{c}{$D\to\rho\ell^+\nu_\ell$} & \multicolumn{2}{c}{$D\to\omega\ell^+\nu_\ell$} & \multicolumn{2}{c}{$D\to K^*\ell^+\nu_\ell$} \\
                                   & This Work                      & CCQM~\cite{Ivanov:2019nqd}& This Work         & CCQM~\cite{Ivanov:2019nqd}& This Work         & CCQM~\cite{Ivanov:2019nqd}        \\ \hline
$\langle P_{\rm L}^e \rangle$            & $+1.000^{+0.168}_{-0.209}$     & $+1.00$     & $+1.000^{+0.188}_{-0.244}$      & $+1.00$      & $+1.000^{+0.136}_{-0.170}$     & $+1.00$     \\
$\langle P_{\rm L}^\mu \rangle$          & $+0.968^{+0.170}_{-0.211}$     & $+0.92$     & $+0.969^{+0.190}_{-0.245}$      & $0.92$       & $+0.958^{+0.138}_{-0.171}$     & $+0.91$     \\
$\langle P_{\rm T}^e \rangle\times 10^2$ & $-0.093^{+0.026}_{-0.114}$     & $-0.09$     & $-0.092^{+0.024}_{-0.018}$      & $-0.09$      & $-0.106^{+0.023}_{-0.019}$     & $-0.11$     \\
$\langle P_{\rm T}^\mu \rangle$          & $-0.189^{+0.042}_{-0.053}$     & $-0.13$     & $-0.186^{+0.048}_{-0.037}$      & $-0.12$      & $-0.213^{+0.046}_{-0.038}$     & $-0.15$     \\
$\langle F_{\rm L}^e\rangle$             & $+0.457^{+0.055}_{-0.067}$     & $+0.53$     & $+0.441^{+0.045}_{-0.057}$      & $+0.52$      & $+0.472^{+0.036}_{-0.042}$     & $+0.54$     \\
$\langle F_{\rm L}^\mu\rangle$           & $+0.461^{+0.053}_{-0.065}$     & $+0.51$     & $+0.445^{+0.044}_{-0.055}$      & $+0.50$      & $+0.478^{+0.035}_{-0.041}$     & $+0.52$   \\
$\langle{\cal A}_{\rm FB}^e\rangle$    & $-0.293^{+0.094}_{-0.117}$     & $-0.21$     & $-0.203^{+0.071}_{-0.094}$      & $-0.21$      & $-0.208^{+0.052}_{-0.066}$     & $-0.18$     \\
$\langle{\cal A}_{\rm FB}^\mu\rangle$  & $-0.279^{+0.091}_{-0.113}$     & $-0.24$     & $-0.191^{+0.068}_{-0.090}$      & $-0.24$      & $-0.192^{+0.049}_{-0.062}$     & $-0.21$     \\
$\langle{\cal C}_{\rm F}^e\rangle$       & $-0.278^{+0.165}_{-0.205}$     & $-0.44$     & $-0.242^{+0.173}_{-0.222}$      & $-0.43$      & $-0.312^{+0.123}_{-0.151}$     & $-0.47$     \\
$\langle{\cal C}_{\rm F}^\mu\rangle$     & $-0.268^{+0.162}_{-0.199}$     & $-0.36$     & $-0.233^{+0.169}_{-0.216}$      & $-0.35$      & $-0.297^{+0.119}_{-0.146}$     & $-0.37$     \\
\hline
\end{tabular}
\addtolength{\arraycolsep}{-20pt}
\label{tab:pol}
\end{table*}

\begin{itemize}
  \item For the top parts, the final lepton polarization for the $D\to V(\rho,\omega,K^*) \ell^+\nu_\ell$ are shown and calculated by applying Eq.~\eqref{Eq:PLTFLT}. All lepton polarizations ($P_{\rm{L,T}}^\ell$) exhibit similar behavior. In the large $q^2$-region, all $P_{\rm{L,T}}^\ell$ are almost unchanged, except that $P_{\rm{T}}^\mu$ rises slowly with the increase of $q^2$, i.e. $P_{\rm{L}}^\ell\approx 1 $, $P_{\rm{T}}^e \approx 0$ and $P_{\rm{T}}^e \lesssim 0$. In the low $q^2$-region, all $P_{\rm{L,T}}^\ell$ polarities are singular due to the $\delta_\ell$ factor, which are clearly shown in the corresponding small graph with the logarithmic axis. We observe that $P_{\rm{L}}^{e(\mu)}$ are approximately equal to $-0.4$ at $q^2_{\rm min}=m_{e(\mu)}^2$. As $q^2$ increases, $P_{\rm{L}}^{e(\mu)}$ then rapidly increases to near $1$ and final remains stable. For transverse component, $P_{\rm{T}}^{e,\mu}(q^2_{\rm min}= m_{e(\mu)}^2) \approx-0.8$. As $q^2$ increases, $P_{\rm{T}}^{\mu}$ rapidly increases to near 0 and then remains stable, while $P_{\rm{T}}^{e}$ increase rapidly and then moderately.
  \item For the bottom parts, the longitudinal $F_{\rm L}^\ell(q^2)$ and transverse $F_{\rm T}^\ell(q^2)$ polarization fractions of the vector meson are shown and calculated by using the Eq.~\eqref{Eq:PLTFLT}, which indicate the three kinds of vector $\rho, \omega, K^*$-mesons have the similar behavior for both $F_{\rm L}^\ell(q^2)$ and $F_{\rm T}^\ell(q^2)$. At all the allowed physical regions, we have $F_{\rm L}^\ell(q^2)+ F_{\rm T}^\ell(q^2)=1$. For the large recoil point $q^2=0~{\rm GeV}^2$, we observe $F_{\rm L}^\ell(0)=1$ and $F_{\rm T}^\ell(0)=0$. As the $q^2$ increases, $F_{\rm L}^\ell(q^2)$ monotonically decreases, and the $F_{\rm T}^\ell(q^2)$ reverses. In addition, $F_{\rm L}^\ell(q^2)$ is dominant the small $q^2$-regions, while $F_{\rm T}(q^2)$ is dominant the large $q^2$-regions. The position of the alternating point of the dominant $q^2$-region is near the midpoint of the whole physically feasible region. At the alternating point $q_{\rm mix}^2$, we observe $F_{\rm L}^\ell=F_{\rm T}^\ell=0.5$ according to the relation $F_{\rm L}^\ell(q^2)+ F_{\rm T}^\ell(q^2)=1$.
\end{itemize}

We then plot the forward-backward asymmetry ${\cal A}_{\rm FB}^\ell$ and the lepton-side ${\cal C}_{\rm F}^\ell(q^2)$ convexity parameter in Fig.~\ref{Fig:AFBCFl}.
\begin{itemize}
 \item The top parts of Fig.~\ref{Fig:AFBCFl} show the change of forward-backward asymmetry ${\cal A}_{\rm FB}^\ell$ from $q_{\rm min}^2=m_\ell^2$ to $q^2_{\rm max}= (m_D-m_V)^2$. All ${\cal A}_{\rm FB}^\ell$ are, first go down from positive value to 0 rapidly, then down to the minimum value slowly, and finally almost level off. ${\cal A}_{\rm FB}^{e(\mu)}=0$ is around $q^2=0.1~{\rm GeV}^2$ ($q^2=10^{-4}~{\rm GeV}^2$), and all ${\cal A}_{\rm FB,max}^\ell\approx0.5$ are shown in the small figure at $q^2=m_e^2$ ($q^2=m_\mu^2$). All of these phenomena can be derived from its analytic expression Eq.~\eqref{eq:AFB}. Therefore, one shall be especially careful when dealing with ${\cal A}_{\rm FB}^\ell$ in the small $q^2$-region, while it will be easier to study in the large $q^2$-region due to the relatively stable value of the ${\cal A}_{\rm FB}^\ell$.
 \item For the bottom parts of Fig.~\ref{Fig:AFBCFl}, we obsreve that ${\cal C}_{\rm F}^\ell(q^2)\leq 0$, ${\cal C}_{\rm F}^\ell (q^2_{\rm min} = m_{e,\mu}^2) = 0$. All the ${\cal C}_{\rm F}^\ell(q^2)$ decrease sharply and then increase, and there is a singularity around low $q^2$, which are exhibited in the small graph with the logarithmic axis.
\end{itemize}

The mean values of those polarization observations of the three $D \to V$ semileptonic decay channels are calculated by applying Eq.~\eqref{eq:Aob} and listed in Table~\ref{tab:pol}. Our predictions are the same as the CCQM results within errors.

\section{Summary}\label{sec:IV}
In this paper, the $D \to V(\omega,\rho, K^*)$ HFFs ${\cal D}_{V,\sigma}$ with $\sigma = 0,1,2,t$ have been studied by applying the LCSR and taking into account the LCDAs up to twist-4. The resultant LCSRs for the HFFs are arranged according to the twist structure of the final vector meson LCDAs. Those HFFs are extrapolated to the whole physics $q^2$-region $m_\ell^2 \leq q^2 \leq (m_D-m_V)^2$, and then we use it to investigate the physical observable for the $D \to V(\rho,\omega,K^*) \ell^+\nu_\ell$ semileptonic decays.

The transversal HFFs and its errors increase as $q^2$ decreases due to the depression effect from $q^2$ coefficient, especially for $\mathcal{D}_{V,1,2}(0)=0 (\pm0)$ with $V=\rho,\omega,K^*$-mesons. This depression effect from $q^2$ coefficient will also be reflected in the different transverse decay width through transversal HFFs, which can be clearly seen from the Fig.~\ref{fig:dGLT}, we also find $1/{|V_{cq}|^2}\times d\Gamma_{V}^{\rm T}(0)=0 (\pm0)$ with $V=\rho,\omega,K^*$-mesons. However, this depression effect for the longitudinal part will disappear due to the missing of $q^2$ coefficient. Thus, the transverse differential decay width dominates the small $q^2$-region, while the longitudinal differential decay width dominates the large $q^2$-region, and the position of the alternating point of the dominant $q^2$-region is near the midpoint of the whole physically feasible region. In addition, the decay width (transverse, longitudinal and total decay width) decreases with the increase of meson mass in final state, and the differences between the transverse decay width and longitudinal decay width are almost the same for the three decay channels, which can be seen from the Table~\ref{tab:GALL}. With the help of lifetime $\tau({D^0})$ and $\tau({D^+})$, we calculate the branching ratio and list it in Table~\ref{tab:BrD2V}. Our predictions are lower compared to other theories, but it fits well with BES-III.

We also investigate in detail the polarization observations dependence on squared momentum transfer for $D \to V(\rho,\omega,K^*) \ell^+\nu_\ell$ semileptonic decays with $\ell=e,\mu$, which has a similar shape for different final mesons and the same final lepton. In the small $q^2$-region, all those polarization observations have a singularity due to the $\delta_\ell$ factor, which are shown in the small graph with the logarithmic axis, expect for $F_{\rm {L, T}}^\ell$. With the increase of $q^2$, all polarization values tend to be more stable, thus the polarization dependence on $q^2$ is declines. Note that $F_{\rm{L}}^\ell$ and $F_{\rm{T}}^\ell$ dominate the small $q^2$-region and the large $q^2$-region respectively, and the position of the alternating point of the dominant $q^2$-region is near the center of the whole physically feasible region, which roughly equates to the positions of the dominant alternating points of the transverse and longitudinal differential decay width. We also calculate the corresponding average values and list them in Table~\ref{tab:pol}, which are coincide with CCQM within the errors.

\section*{Acknowledgments}
We are grateful to Prof. Xing-Gang Wu for helpful communications and discussions. This work was supported in part by the National Natural Science Foundation of China under Grant No.11765007, 11947302, 11947406, the Project of Guizhou Provincial Department of Science and Technology under Grant No.KY[2019]1171, the China Postdoctoral Science Foundation under Grant No.2019TQ0329, the Project of Guizhou Minzu University under Grant No. GZMU[2019]YB19. This work was done during stay@home international action to stop the COVID-19 epidemic.

\section*{Appendix}
\begin{table}[t]
\centering
\small
\caption{The moments and couplings of vector meson twist-2,3,4 LCDA, the corresponding scale are $\mu^2=m_D^2-m_c^2\approx1 \rm{GeV}^2$.}
\renewcommand{\arraystretch}{1.5}
\addtolength{\arraycolsep}{20pt}
\begin{tabular}{cccc}\\[-2ex]
\hline
                 & $\rho$     & $\omega$   & $K^*$     \\ \hline
$a_1^\|$         & 0          & 0          & $0.19(5)$ \\
$a_2^\|$         & $0.18(10)$ & $0.18(10)$ & $0.06(6)$ \\
$a_1^\bot$       & 0          & 0          & $0.20(5)$ \\
$a_2^\bot$       & $0.20(10)$ & $0.18(10)$ & $0.04(4)$ \\
$\delta_+$       & 0          & 0          & $0.24$    \\
$\delta_-$       & 0          & 0          & $-0.24$   \\
$\tilde\delta_+$ & 0          & 0          & $0.16$    \\
$\tilde\delta_-$ & 0          & 0          & $-0.16$   \\ \hline
\end{tabular}
\addtolength{\arraycolsep}{-20pt}
\label{tab:moment}
\end{table}
In order to get the accurate HFFs results within LCSR approach for the semileptonic decay processes $D\to V(\rho,\omega,K^*)\ell^+\nu_\ell$ and make a comparison with other theoretical and experimental results, we take the twist-2,3,4 LCDAs given by P. Ball and V. M. Braun~\cite{Ball:1998kk} used by many theoretical predictions. The two-particle LCDAs for twist-3 has the following form,
\begin{align}
\psi_{3;V}^\bot(u) & =  6 u \bar u \bigg[1 + a_1^\parallel \xi + \bigg\{\frac{1}{4}a_2^\parallel + \frac{5}{3} \zeta_{3} \bigg(1-\frac{3}{16}\omega^A_{3}\nonumber\\
& +\frac{9}{16}\omega^V_3\bigg)\bigg\}(5\xi^2-1)\bigg]+ 6 \tilde{\delta}_+  (3u \bar u + \bar u \ln \bar u
\nonumber\\
& + u \ln u) + 6 \tilde{\delta}_- (\bar u \ln \bar u - u \ln u),
\\[2ex]
\phi_{3;V}^\bot (u) & = \frac{3}{4}(1+\xi^2) + a_1^\| \frac{3}{2}\xi^3 + \bigg(\frac{3}{7} a_2^\|+ 5 \zeta_3 \bigg) \nonumber\\
&\times(3\xi^2-1)  + \left[ \frac{9}{112} a_2^\|
+ \frac{15}{64}\zeta_3\Big(3\omega_3^V-\omega_3^A\Big)
 \right]
\nonumber\\
&\times (3-30\xi^2 + 35\xi^4) +\frac32 \widetilde{\delta}_+(2+\ln u + \ln\bar u)
\nonumber\\
&+\frac32\widetilde{\delta}_-\, ( 2 \xi + \ln\bar u - \ln u),
\label{eq:gv}
\\[2ex]
\psi_{3;V}^\|(u) & = 6u\bar u \left[ 1 + a_1^\perp \xi + \left( \frac{1}{4}a_2^\perp + \frac{5}{8}\zeta_{3}\omega_3^T \right) (5\xi^2-1)\right]
\nonumber\\
&+ 3 \delta_+ (3 u \bar u + \bar u \ln \bar u + u \ln u) + 3\delta_-  (\bar u
 \ln \bar u - u
\nonumber\\
& \times\ln u),
\label{eq:e}
\\[2ex]
\phi_{3;V}^\| (u) &=  3\xi^2 + \frac{3}{2}a_1^\perp \xi (3 \xi^2-1)
+ \frac{3}{2} a_2^\perp \xi^2 (5\xi^2-3)
\nonumber\\
&+ \frac{15}{16}\zeta_{3}\omega_3^T(3 - 30\xi^2 + 35\xi^4) + \frac{3}{2} \delta_+ \nonumber\\
& \times\bigg(1 + \xi\ln\frac{\bar u}{u}\bigg) + \frac{3}{2}\delta_-  \xi ( 2 + \ln u + \ln\bar u).
\label{eq:hL}
\end{align}
And the two-particle LCDAs for twist-4 can be written as
\begin{align}
\psi_{4;V}^\bot(u) &= 6u(1-u)+5\left[\zeta^{\rm T}_4+\tilde \zeta^{\rm T}_4\right](1-3\xi^2),
\\[2ex]
\phi_{4;V}^\bot(u) &= 30 u^2(1-u)^2 \Bigg[\frac{2}{5}+\frac{4}{3}\zeta^{\rm T}_4-\frac{8}{3}\tilde \zeta^{\rm T}_4\Bigg],
\\[2ex]
\phi_{4;V}^\|(u)   &= \Bigg[\frac{4}{5}+\frac{20}{9} \zeta_4
                    +\frac{8}{9} \zeta_3\Bigg]30 u^2\bar u^2,
\\
\psi_{4;V}^\|(u)   &= 6u\bar u + \Bigg[\frac{10}{3} \zeta_4
                    -\frac{20}{3} \zeta_3\Bigg](1-3 \xi^2),
\\
C_V(u) &= \Bigg[\frac{3}{2}+\frac{10}{3} \zeta_{4}
                    +\frac{10}{3} \zeta_3\Bigg]u^2\bar u^2,
\end{align}
The values of the moments and coupling constants of the vector meson twist-2,3,4 LCDAs are listed in Table~\ref{tab:moment}. At the scale $\mu^2=m_D^2-m_c^2\approx1 \rm{GeV}^2$, the couplings for twist-3 and twist-4 LCDAs are
\begin{align}
&\zeta_3 = 0.032,~~\omega_3^A = - 2.1,~~ \omega_3^V = 3.8,~~ \omega_3^T = 7.0,\nonumber\\
&\zeta_4=0.15,~~\zeta_4^T = 0.10,~~ \tilde\zeta_4^T = -0.10.
\end{align}

\end{document}